\documentclass[aps,prc,twocolumn,groupedaddress,amsmath,amssymb,showpacs,showkeys,nofootinbib]{revtex4-2}
\usepackage{graphicx}
\usepackage{dcolumn}
\usepackage{bm}
\usepackage{threeparttable}
\usepackage{enumitem}

\begin{document}


\title{Search for the $\alpha$ + core structure in the ground state bands of \linebreak[4] $22 \leq Z \leq 42$ even-even nuclei}

\author{M.~A.~Souza}
\email{marsouza@if.usp.br}
\affiliation{Instituto de F\'{\i}sica, Universidade de S\~{a}o Paulo, Rua do Mat\~{a}o, 1371, CEP 05508-090, Cidade Universit\'{a}ria,
S\~{a}o Paulo - SP, Brazil}
\author{H.~Miyake}
\email{miyake@if.usp.br}
\affiliation{Instituto de F\'{\i}sica, Universidade de S\~{a}o Paulo, Rua do Mat\~{a}o, 1371, CEP 05508-090, Cidade Universit\'{a}ria,
S\~{a}o Paulo - SP, Brazil}

\date{\today}

\begin{abstract}
A systematic analysis of the $\alpha$ + core structure is performed in the ground state bands of even-even nuclei of the \mbox{$22 \leq Z \leq 42$} region in terms of the Local Potential Model. The $\alpha$ + core interaction is described by the nuclear potential of \mbox{(1 + Gaussian)$\times$(W.S.~+ W.S.$^3$)} shape with 2 free parameters. Properties such as energy levels, reduced $\alpha$-widths, $B(E2)$ transition rates and rms charge radii are calculated and compared with experimental data. A good agreement with the experimental data is obtained in general, even for the nuclei without the \mbox{$\alpha$ + \{doubly closed shell core\}} configuration. The analysis of the selected nuclei in the \mbox{$22 \leq Z \leq 42$} region indicates that $^{44}$Ti and $^{94}$Mo have a greater $\alpha$-clustering degree in comparison with their respective neighboring nuclei of this set. In addition, the model points to the existence of nuclei without the \mbox{$\alpha$ + \{doubly closed shell core\}} configuration with a significant $\alpha$-clustering degree compared to $^{44}$Ti, $^{60}$Zn, and $^{94}$Mo. The study shows that the $\alpha$ + core approach is satisfactorily applicable to nuclei other than those with $\alpha$-clustering above double-shell closures.
\end{abstract}

                            
                              
\maketitle

\section{Introduction}

Cluster models constitute a topic of nuclear physics in continuous development, being used for the investigation of properties in nuclei of different mass regions. The most recent challenges in the light mass region, such as the description of the Hoyle state, Hoyle-like states, and $\alpha$ gas states in general, are discussed by P.~Schuck in Ref.~\cite{S2018}, and the techniques applied in heavier nuclei, such as the antisymmetrized molecular dynamics, the quartetting wave function approach, etc.~are detailed by Z.~Ren and B.~Zhou in Ref.~\cite{RZ2018}. In the experimental field, an overview of the recent results in nuclear cluster physics is discussed in detail by E.~Vardaci in Ref.~\cite{V2018}.

An approach widely used in nuclear cluster physics is the Local Potential Model (LPM), as presented originally by Buck, Dover and Vary \cite{BDV75}. In this approach, the nucleus is assumed as a cluster + core system where the two components interact through a deep local potential $V(r)$ containing the nuclear and Coulomb terms; in the case where cluster and/or core has non-zero spin, terms resulting from non-central forces are added such as the spin-orbit term. In the last decades, $\alpha$-clustering has been analyzed by LPM in several studies emphasizing the nuclei with the $\alpha$ + \{doubly closed shell core\} (hereafter: $\alpha + \mathrm{DCSC}$) configuration, such as $^{20}$Ne, $^{44}$Ti, $^{52}$Ti, $^{60}$Zn, $^{94}$Mo, and $^{212}$Po (examples in Refs.~\cite{BMP95,MRO88,O1995,MOR1998,BJM1995,BMP1999_PRC61,MRO2000,SM2015,HMS1994,M2008,WPX2013,O2020,SMB2019,IMP2019}). In this way, some works demonstrate that the nuclei with the $\alpha + \mathrm{DCSC}$ configuration can be systematically analyzed with the same $\alpha$ + core potential type, providing a good general agreement with experimental data, as shown in Refs.~\cite{BMP95,WPX2013,SMB2019,IMP2019,BR2021}. In the recent work of D.~Bai and Z.~Ren \cite{BR2021}, a study on the $\alpha$-cluster structure above double shell closures via double-folding potentials is shown, analyzing the $^{8}$Be, $^{20}$Ne, $^{44,52}$Ti, and $^{212}$Po nuclei; this study presents a detailed connection of the $\alpha$ + core potential with the chiral effective field theory.

Due to experimental data published in 2018 by Auranen {\it et al.}~on $^{104}$Te \cite{ASA2018}, further work was made on $\alpha$-clustering above the double shell closure at $^{100}$Sn, such as D.~Bai and Z.~Ren \cite{BR2018}, M.~A.~Souza {\it et al.}~\cite{SMB2019}, and T.~T.~Ibrahim {\it et al.}~\cite{IMP2019}, including the microscopic calculation of S.~Yang {\it et al.}~\cite{YXR2020}, obtaining results compatible with the experimental $\alpha$-decay half-life of $^{104}$Te. Additional experimental data on energy levels, electromagnetic transition rates, half-lives of the excited states, etc.~are needed to better evaluate the theoretical predictions on $^{104}$Te.

In recent years, the Ti isotopes have been investigated theoretically and experimentally from the perspective of the $\alpha$-cluster structure \cite{O2020,BKF2019,BKF2021,NUN2020}. In the studies of S.~Bailey {\it et al.}~\cite{BKF2019,BKF2021}, a novel technique makes use of the continuous wavelet transform and machine learning to identify $\alpha$-clustered states in $^{44,48,52}$Ti through \mbox{$^4$He($^{40,44,48}$Ca, $\alpha$)} resonant scattering measurements; such studies indicate the presence of the $\alpha + ^{40}$Ca and $\alpha + ^{48}$Ca structures in high-lying states of $^{44}$Ti and $^{52}$Ti, respectively. The results of S.~Bailey {\it et al.}~reinforce the indications of previous experiments which also point to the presence of the $\alpha$-cluster structure in Ti isotopes, as the $^{40,42,48}$Ca($^6$Li, $d$)$^{44,46,52}$Ti \cite{YOF1990,GJZ1993,YII1996,YKF1998,FBL1977} and $^{40,42}$Ca($^7$Li, $t\alpha$)$^{40,42}$Ca \cite{FTO2009} reactions, and other $\alpha$-transfer processes on Ca targets. The work of S.~Ohkubo \cite{O2020} discusses the $\alpha + ^{48}$Ca structure in $^{52}$Ti, using the optical potential model to analyze $\alpha + ^{48}$Ca scattering data; the real part of the optical potential was used to calculate the $\alpha + ^{48}$Ca energy bands, indicating that three $^{52}$Ti experimental states found by S.~Bailey {\it et al.}~\cite{BKF2019} are associated with the higher nodal positive parity band of the $\alpha + ^{48}$Ca system.

About the Cr isotopic chain, the present authors investigated the $\alpha$ + core structure in $^{46}$Cr and $^{54}$Cr through LPM \cite{SM2017}, introducing the use of the \mbox{(1 + Gaussian)$\times$(W.S.~+ W.S.$^3$)} nuclear potential; the model showed good agreement with the energies of the ground state band and experimental $B(E2)$ transition rates, and it was shown that $^{46}$Cr has a significant $\alpha$-clustering degree compared to the well-studied $^{44}$Ti. A similar study of P.~Mohr \cite{M2017} on $^{46}$Cr and $^{54}$Cr also showed satisfactory results for the same properties using a double-folding nuclear potential with the DDM3Y interaction. Previous works by Descouvemont \cite{D2002} and Sakuda and Ohkubo \cite{SO2002} describe $^{48}$Cr in terms of an $\alpha + \alpha + ^{40}$Ca system using the generator coordinate method and orthogonality condition model, respectively, obtaining good results in general for energy levels and $B(E2)$ rates.

A natural path in the development of the $\alpha$-cluster model was its application in the $A \sim 90$ region, due to the proximity of the neutron shell closure at $N = 50$ and proton subshell closure at $Z = 40$. Calculations on $^{94}$Mo in terms of an $\alpha + ^{90}$Zr system were made by different authors with favorable results \cite{BMP95,O1995,MOR1998,MRO2000,SM2015,M2008,SMB2019,IMP2019}. In order to extend the application of the model to nuclei around $^{94}$Mo, the present authors made a study on the \mbox{$\alpha$ + core} structure in $^{90}$Sr, $^{92}$Zr, $^{94}$Mo, $^{96}$Ru, and $^{98}$Pd using the \mbox{W.S.~+ W.S.$^3$} nuclear potential \cite{SM2015}, showing that LPM produces a good general description of the energy spectra, $B(E2)$ transition rates, and rms charge radii, suggesting a systematic behavior in their \mbox{$\alpha$ + core} structures. The previous work of P.~Mohr \cite{M2008} shows a similar treatment of the $\alpha$ + core system involving nuclei with $N = 52$ ($N_{\mathrm{core}} = 50$) from $^{82}$Zn to $^{104}$Te, producing satisfactory results for the experimental $B(E2)$ rates without the use of effective charges; the double-folding nuclear potential used in Ref.~\cite{M2008} requires a strength parameter $\lambda$ with a smooth dependence on the quantum number $L$ to reproduce the experimental ground state bands.

Our studies in Refs.~\cite{SM2015,SM2017,SMB2019} suggest that LPM can be applied systematically in nuclei of distinct mass regions, including cases without the \mbox{$\alpha + \mathrm{DCSC}$} configuration. Therefore, the present work proposes a comprehensive study of the \mbox{$\alpha$ + core} structure in several even-even nuclei in the $Z$-region delimited by two nuclei with a well-recognized $\alpha$-cluster structure: $^{44}$Ti and $^{94}$Mo ($22 \leq Z \leq 42$ or $20 \leq Z_{\mathrm{core}} \leq 40$). For this, the \mbox{(1 + Gaussian)$\times$(W.S.~+ W.S.$^3$)} nuclear potential, with 2 free parameters and 4 fixed parameters, and already used successfully in Refs.~\cite{SM2017,SMB2019}, is applied in the present work. A comparative study of the nuclei selected in this region is carried out to identify those with higher degree of $\alpha$-clustering.

\section{Selection of nuclei for analysis}
\label{Sec:Q/A}

As exemplified in the previous section, a criterion applied commonly in studies on the $\alpha + \mathrm{core}$ structure is the selection of nuclei with doubly or simply magic core. In the present study, we use the same criterion applied in our previous works \cite{SM2015, SM2017} where the different even-$Z$ isotopic chains are analyzed through the variation of binding energy per nucleon due to the $\alpha$-core separation

\begin{equation}
\frac {Q_\alpha}{A_T}=\frac{B_\alpha +B_{\mathrm{core}}-B_T}{A_T}\;, 
\end{equation}

\noindent where $Q_\alpha$ is the $Q$-value for $\alpha$-separation, $A_T$ is the mass number of the total nucleus and $B_\alpha $, $B_{\mathrm{core}}$ and $B_T$ are the experimental binding energies of the $\alpha $-cluster, core and total nucleus, respectively. An absolute (or local) maximum of $Q_{\alpha}/A_T$ is considered to indicate the preferential nucleus for $\alpha$-clustering compared to the remaining (or neighboring) nuclei of the set. The values of $B_\alpha $, $B_{\mathrm{core}}$ and $B_T$ are taken from Ref.~\cite{HAW2017}.

There are nuclei in the $22 \leq Z \leq 42$ region with unmeasured spectra, especially unstable nuclei. Therefore, it is possible that the criterion described above selects nuclei in which there is insufficient experimental data for a consistent analysis of the $\alpha + \mathrm{core}$ structure. Taking this possibility into account, a first additional condition is established for the selection of nuclei in the isotopic chain: {\it (i)} if the highest $Q_{\alpha}/A_T$ nucleus does not have at least some experimental energy levels with fairly defined spins and parities, one selects the isotope with the $Q_{\alpha}/A_T$ value immediately below which has more experimental energy levels with defined assignments. Also, a second condition is defined for an analysis of the influence of the shell closures on the $\alpha$-clustering degree: {\it (ii)} the isotopes with doubly or simply magic core should be selected for a comparative study, provided they have at least some experimental energy levels with fairly defined spins and parities. The only exception to the selection criterion established is the inclusion of $^{48}$Cr in the study, for the reason explained below.

\begin{figure*}
\includegraphics[scale=0.8]{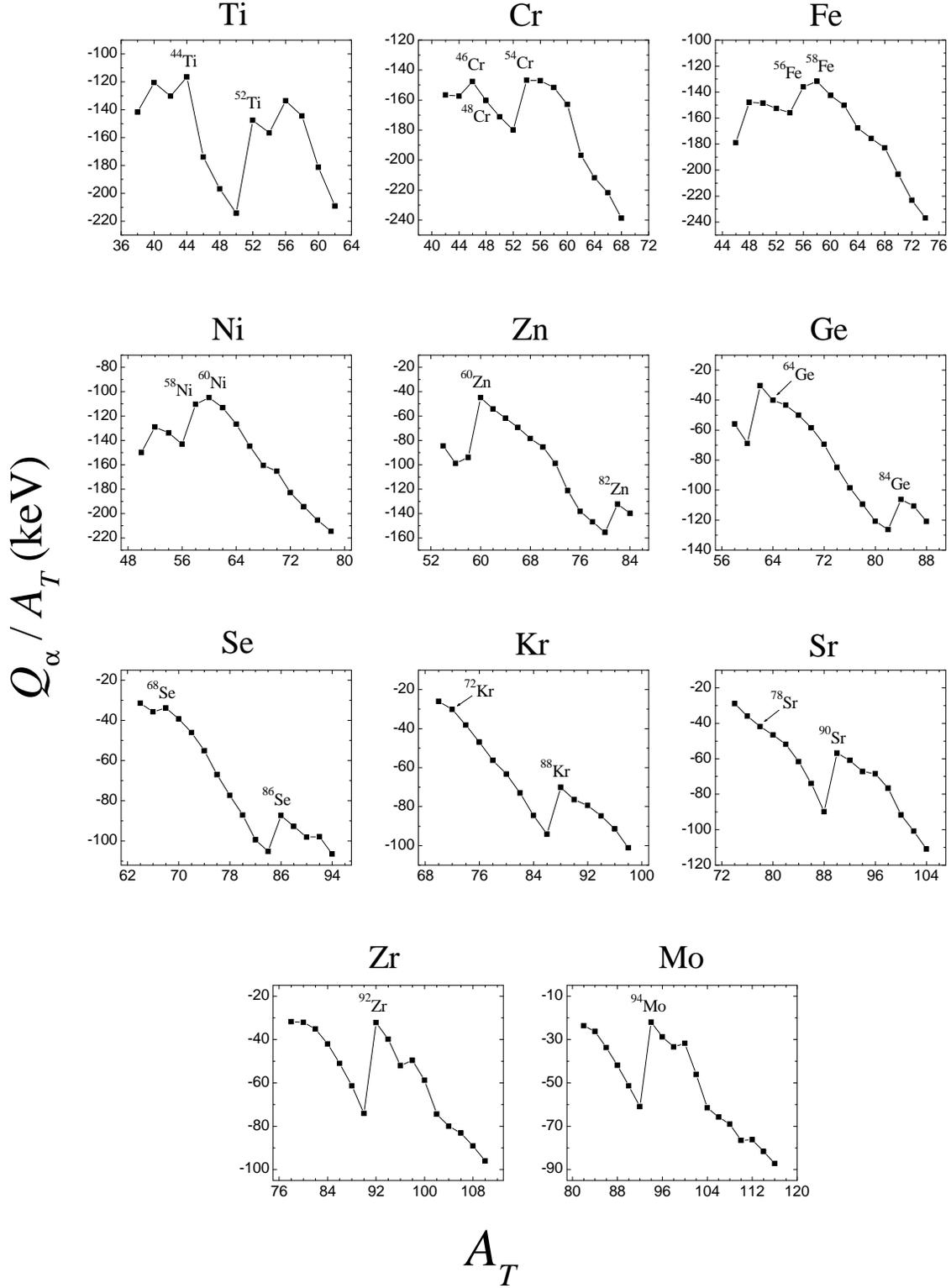}
\caption{$Q_{\alpha}/A_T$ values obtained for the $\alpha$ + core decomposition of even-even nuclei in the isotopic chains from Ti ($Z_T = 22$) to Mo ($Z_T = 42$) as a function of the total mass number $A_T$. The nuclei selected for the $\alpha$-cluster analysis are indicated.}
\label{Fig_QdivA}
\end{figure*}

Fig.~\ref{Fig_QdivA} shows graphically the $Q_{\alpha}/A_T$ values obtained for the even-even nuclei of the isotopic chains from Ti to Mo, indicating the nuclei selected for analysis. The selection of nuclei in each isotopic chain is detailed below:

\begin{enumerate}[label={},leftmargin=0.2cm,align=left]
\item {\bf Ti, Fe, Ni, and Zn}: the $^{44}$Ti, $^{58}$Fe, $^{60}$Ni, and $^{60}$Zn nuclei are those with the highest $Q_{\alpha}/A_T$ values in their respective isotopic chains, being selected for analysis. As the $^{52}$Ti, $^{56}$Fe, $^{58}$Ni and $^{82}$Zn nuclei have doubly or simply magic core, they were selected for a comparative study.

\item {\bf Cr}: this isotopic chain was discussed in our previous work on $^{46,54}$Cr \cite{SM2017}. $^{54}$Cr has the highest $Q_{\alpha}/A_T$ value in the set of even-even isotopes ($-146.8$ keV). However, $^{46}$Cr has a very close $Q_{\alpha}/A_T$ value ($-147.7$ keV) and has a magic $N_{\mathrm{core}}$ = 20. In this case, both were selected for analysis. As $^{48}$Cr is pointed as a favorable nucleus for $\alpha$-clustering in Refs.~\cite{D2002,SO2002}, it was included in the comparative study exceptionally.

\item {\bf Ge, Se, Kr, and Sr}: the nuclei $^{62}$Ge, $^{64}$Se, $^{70}$Kr, and $^{74}$Sr are those with the highest $Q_{\alpha}/A_T$ values in their respective isotopic chains. However, such nuclei have only the $0^{+}$ ground state known experimentally, or the states above $0^{+}$ have no defined spin and parity. For this reason, the neighboring isotopes with closest $Q_{\alpha}/A_T$ values and identified experimental levels were selected: $^{64}$Ge, $^{68}$Se, $^{72}$Kr, and $^{78}$Sr. The isotopes with magic $N_{\mathrm{core}} = 50$ were also selected for a comparative study: $^{84}$Ge, $^{86}$Se, $^{88}$Kr, and $^{90}$Sr.

\item {\bf Zr}: The three highest $Q_{\alpha}/A_T$ values in the set of even Zr isotopes are: $-31.825$ keV, $-32.072$ keV, and $-32.200$ keV for $^{78}$Zr, $^{80}$Zr, and $^{92}$Zr, respectively, according to the AME2016 binding energy data \cite{HAW2017}. Such values would lead to the selection of $^{78}$Zr for analysis, however, using the binding energies from the AME2012 table \cite{AWW2012}, it is obtained that $^{92}$Zr is the even isotope with highest $Q_{\alpha}/A_T$. Due to the mentioned discordant results, and taking into account that $^{78}$Zr and $^{80}$Zr have only the $0^{+}$ ground state with defined spin and parity, only $^{92}$Zr was selected in this isotopic chain.

\item {\bf Mo}: The $^{94}$Mo nucleus is the one with the highest $Q_{\alpha}/A_T$ value in the set of even Mo isotopes. There are no other experimentally identified Mo isotopes with magic core. Therefore, only $^{94}$Mo was selected.
\end{enumerate}

\noindent Thus, a total of 21 even-even nuclei were selected for a systematic analysis of the $\alpha + \mathrm{core}$ structure in the $22 \leq Z \leq 42$ region. The criterion defined by the absolute maximum $Q_{\alpha}/A_T$ value naturally selects the $^{44}$Ti, $^{60}$Zn, and $^{94}$Mo nuclei with \mbox{$\alpha + \mathrm{DCSC}$} configuration, while the $^{52}$Ti and $^{82}$Zn nuclei, also with the \mbox{$\alpha + \mathrm{DCSC}$} configuration, were selected by condition {\it (ii)}.

The selected nuclei allow a broad study of the influence of the $Q_{\alpha}/A_T$ value and the magic core condition on the $\alpha$-clustering degree, as seen in Sec.~\ref{Sec:Results}.

\section{$\alpha$-cluster model}

The states of the total nucleus are described in terms of an $\alpha$-particle orbiting an inert core. The $\alpha + \mathrm{core}$ interaction is described by the local potential

\begin{equation}
V(r)=V_N(r)+V_C(r)
\end{equation}

\noindent with the nuclear and Coulomb terms. The Coulomb potential $V_C(r)$ is that of an $\alpha$-particle interacting with an uniformly charged spherical core of radius $R$. The intercluster nuclear potential $V_N(r)$ is given by the expression

\begin{eqnarray}
V_N(r) = -V_{0}\left[1+\lambda\exp\left(-\frac{r^{2}}{\sigma^{2}}\right)\right]
 \left\{ \frac b{1+\exp [(r-R)/a]} \right. {}
                                                          \nonumber\\[4pt]
 {} \left. + \frac{1-b}{\{1+\exp[(r-R)/3a]\}^3}\right\} \;, \qquad \qquad
\label{eq:Nuc_Pot}
\end{eqnarray}

\noindent where $V_0$, $\lambda$, $a$, and $b$ are fixed parameters, and $R$ and $\sigma$ are variable parameters. The \mbox{(1 + Gaussian)$\times$(W.S.~+ W.S.$^3$)} potential shape of eq.~\eqref{eq:Nuc_Pot} has already been successfully applied in our previous works on the $\alpha + \mathrm{core}$ structure in $^{46,54}$Cr \cite{SM2017}, in the set \{$^{20}$Ne, $^{44}$Ti, $^{94}$Mo, $^{212}$Po\} \cite{SMB2019}, and produced \mbox{$Q_{\alpha}$-values} and $\alpha$-decay half-lives for $^{104}$Te in agreement with experimental data \cite{SMB2019} using the same fixed parameters.

The parameter values used are: $V_0 = 220$ MeV, $a = 0.65$ fm, $b = 0.3$, and $\lambda = 0.14$, while $R$ and $\sigma$ are fitted specifically for each nucleus. Details on the origin of the parameter values, as well as the development of the \mbox{(1 + Gaussian)$\times$(W.S.~+ W.S.$^3$)} shape, are explained in Refs.~\cite{SMB2019, SM2017}. The variable parameters $\sigma$ and $R$, shown in Table \ref{Table_parameters}, are fitted to provide the best possible reproduction of the $0^{+}$ and $4^{+}$ experimental levels of the ground state band.

The four nucleons of the $\alpha$-cluster must lie in shell-model orbitals outside those occupied by the core nucleons. This restriction is defined by the global quantum number $G = 2N + L$, where $N$ is the number of internal nodes in the radial wave function and $L$ is the orbital angular momentum. The quantum number $G_{\mathrm{g.s.}}$ associated with the ground state band is shown in Table \ref{Table_parameters} for the selected nuclei. In accordance with the \mbox{Wildermuth} condition \cite{WT1977}, $G_{\mathrm{g.s.}} = 12$, 14, and 16 correspond to the $(pf)^4$, $(pf)^2(sdg)^2$, and $(sdg)^4$ configurations for the valence nucleons, respectively. The valence nucleons of $^{72}$Kr and $^{78}$Sr are in a transition region from the $pf$-shell to the $sdg$-shell; therefore, two numbers $G_{\mathrm{g.s.}} = 12$ and 14 are tested to verify which is the most suitable in describing the g.s.~bands of $^{72}$Kr and $^{78}$Sr, as discussed in more detail in Sec.~\ref{Sec:Results}.

The resolution of the Schr\"{o}dinger radial equation for the $\alpha$ + core relative motion allows to determine the energy levels of the system and respective radial wave functions, which are used to calculate other properties such as rms radii, electromagnetic transition rates, reduced $\alpha$-widths, etc.

\begin{table}
\caption{Values of the parameters $R$ and $\sigma$ and the quantum number $G_{\mathrm{g.s.}}$ for the nuclei studied.}
\label{Table_parameters}
\begin{ruledtabular}
\begin{tabular}{cccc}
Nucleus & $G_{\mathrm{g.s.}}$  & $R$ (fm) & $\sigma$ (fm) \\[2pt] \hline
&  &  &  \\[-6pt]
$^{44}$Ti  &  12   &  4.551  &  0.425 \\
$^{52}$Ti  &  12   &  4.612  &  0.382 \\
$^{46}$Cr  &  12   &  4.658  &  0.248 \\
$^{48}$Cr  &  12   &  4.684  &  0.215 \\
$^{54}$Cr  &  12   &  4.674  &  0.210 \\
$^{56}$Fe  &  12   &  4.694  &  0.303 \\
$^{58}$Fe  &  12   &  4.690  &  0.307 \\
$^{58}$Ni  &  12   &  4.680  &  0.510 \\
$^{60}$Ni  &  12   &  4.670  &  0.545 \\
$^{60}$Zn  &  12   &  4.611  &  0.320 \\
$^{82}$Zn  &  14   &  5.367  &  0.204 \\
$^{64}$Ge  &  12   &  4.647  &  0.278 \\
$^{84}$Ge  &  14   &  5.339  &  0.236 \\
$^{68}$Se  &  12   &  4.678  &  0.247 \\
$^{86}$Se  &  14   &  5.326  &  0.289 \\
$^{72}$Kr  &  12   &  4.724  &  0.000 \\
           &  14   &  5.221  &  0.160 \\
$^{88}$Kr  &  14   &  5.318  &  0.314 \\
$^{78}$Sr  &  12   &  4.793  &  0.000 \\
           &  14   &  5.294  &  0.000 \\
$^{90}$Sr  &  14   &  5.321  &  0.318 \\
$^{92}$Zr  &  14   &  5.295  &  0.248 \\
$^{94}$Mo  &  16   &  5.783  &  0.410 \\
\end{tabular}
\end{ruledtabular}
\end{table}

\section{Results}
\label{Sec:Results}

\begin{figure*}
\includegraphics[scale=0.8]{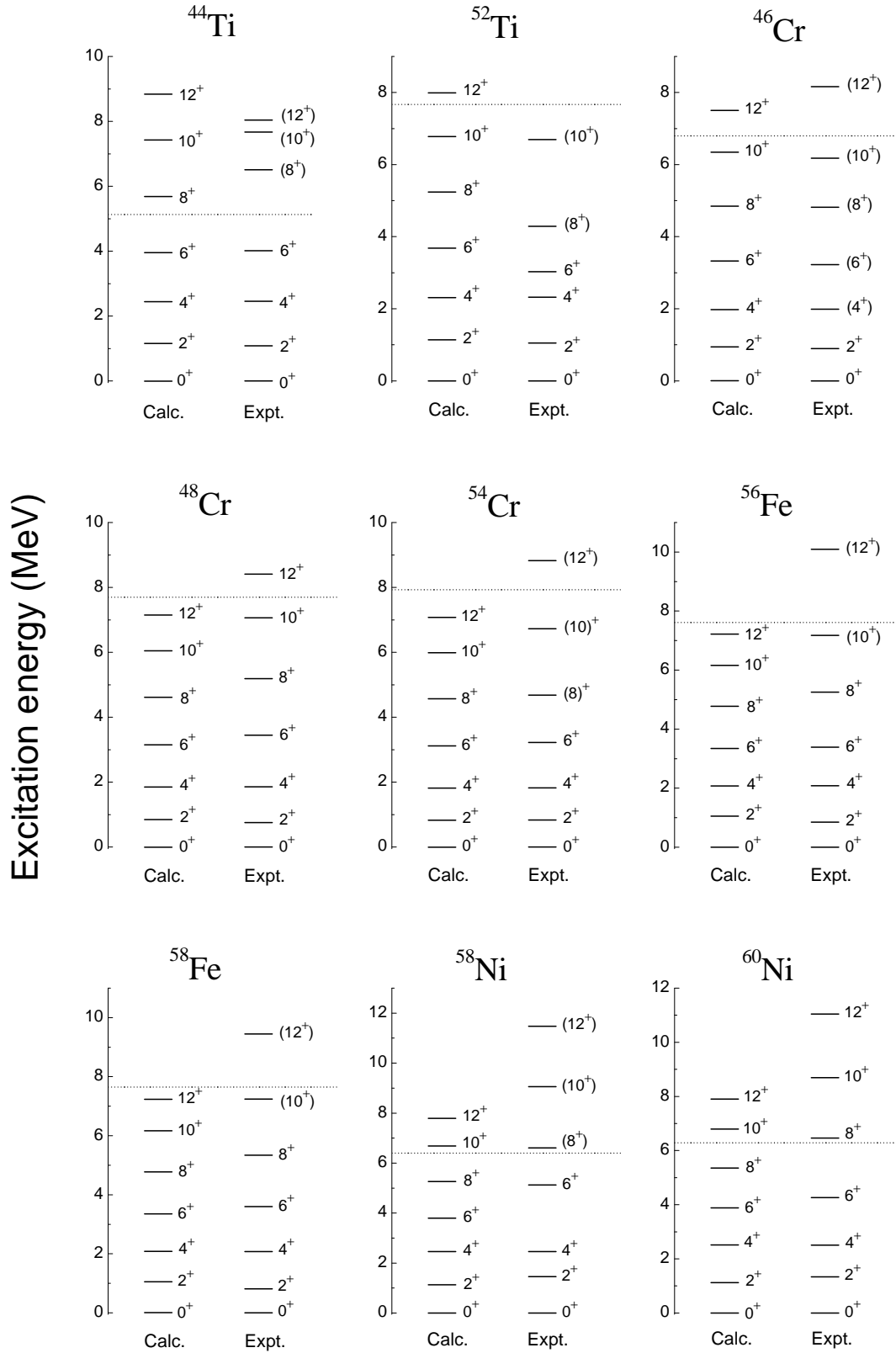}
\caption{Calculated energy levels for the ground state bands of the Ti, Cr, Fe, and Ni isotopes selected in this work in comparison with experimental energies. The dotted lines indicate the $\alpha$ + core thresholds.}
\label{Fig_espec_Ti_Ni}
\end{figure*}

\begin{figure*}
\includegraphics[scale=0.8]{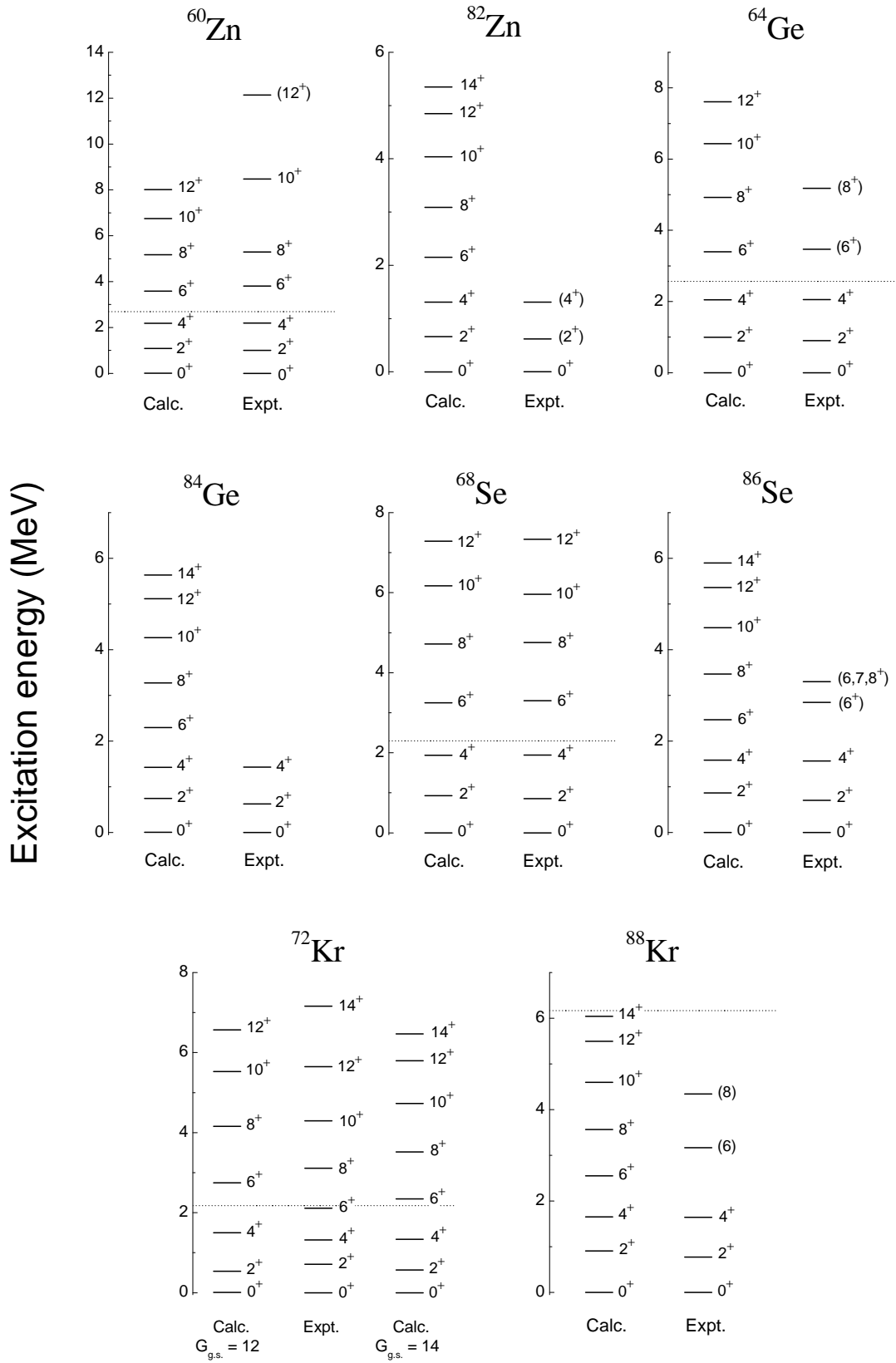}
\caption{Calculated energy levels for the ground state bands of the Zn, Ge, Se, and Kr isotopes selected in this work in comparison with experimental energies. The dotted lines indicate the $\alpha$ + core thresholds. In the cases of $^{82}$Zn, $^{84}$Ge, and $^{86}$Se, the $\alpha$ + core thresholds are $E_{\mathrm{thr.}} = 10.8493$ MeV, 8.9247 MeV and 7.5130 MeV, respectively.}
\label{Fig_espec_Zn_Kr}
\end{figure*}

\begin{figure*}
\includegraphics[scale=0.8]{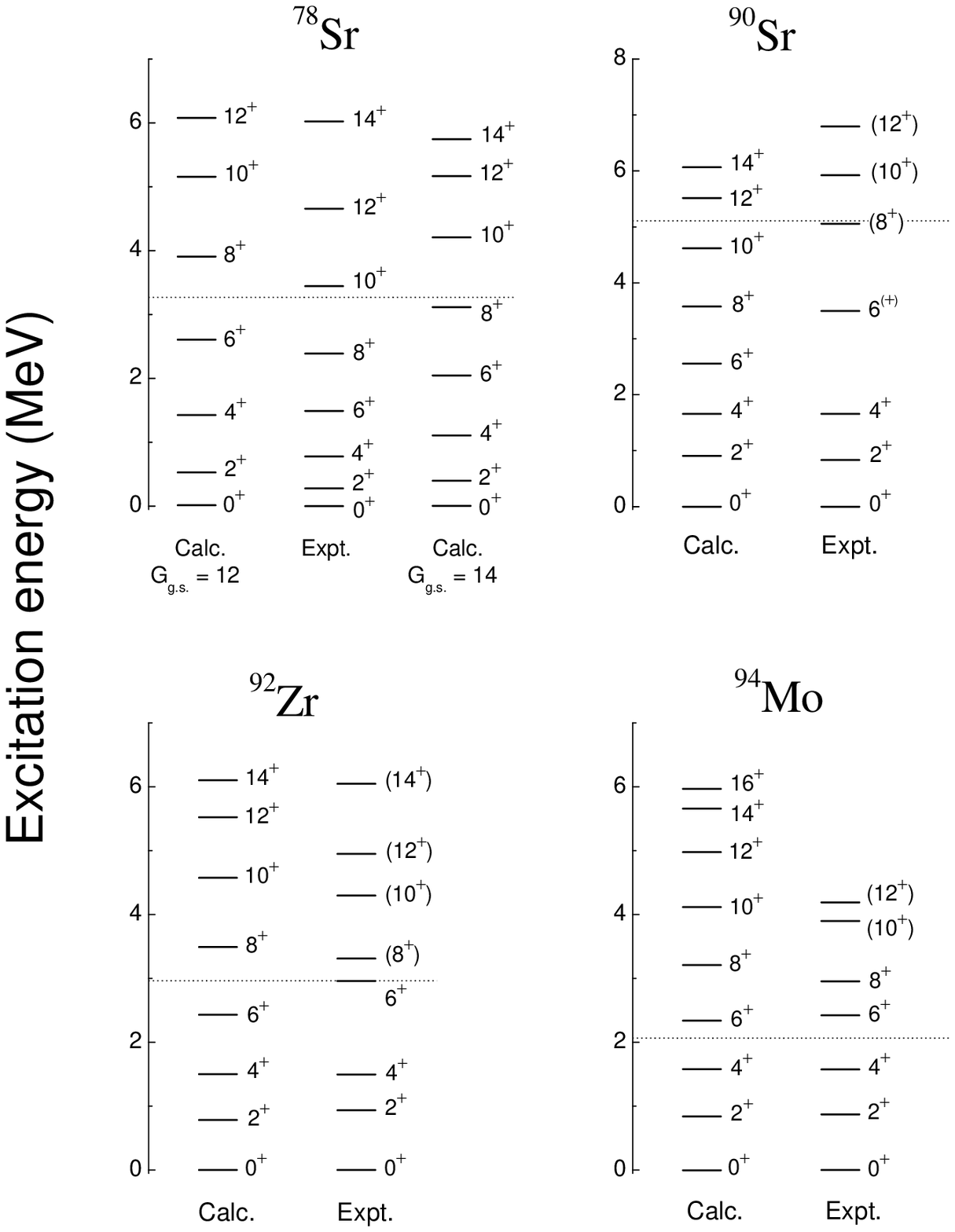}
\caption{Calculated energy levels for the ground state bands of the Sr, Zr, and Mo isotopes selected in this work in comparison with experimental energies. The dotted lines indicate the $\alpha$ + core thresholds.}
\label{Fig_espec_Sr_Mo}
\end{figure*}

The calculated ground state bands are shown in Figs.~\ref{Fig_espec_Ti_Ni}, \ref{Fig_espec_Zn_Kr}, and \ref{Fig_espec_Sr_Mo} in comparison with experimental energies (\cite{ENSDF}, except \cite{DVM2018} for $^{84}$Ge). In general, the proposed $\alpha + \mathrm{core}$ potential provides a satisfactory description of the experimental bands, mainly from $0^{+}$ to $8^{+}$. For the levels from $10^{+}$ to $14^{+}$, the calculated bands for $^{44,52}$Ti, $^{46}$Cr, $^{68}$Se, $^{72}$Kr (with $G_{\mathrm{g.s.}} = 14$), and $^{92}$Zr provide at least reasonable results. The \mbox{(1 + Gaussian)$\times$(W.S.~+ W.S.$^3$)} potential generates a more compressed spacing at the highest spin levels, which differs more strongly from the corresponding experimental levels of some nuclei. However, most experimental levels above $8^{+}$ have uncertain spin and/or parity, or have not been measured, which prevents a more accurate judgment of the high spin levels. 

For a better evaluation of the theoretical spectra, the standard deviation ($S.D.$) of the experimental energy levels in relation to the calculated energy levels is determined (see Table \ref{Table_errors}), taking into account $n - p$ degrees of freedom, where $n$ is the number of energy levels considered, and $p = 2$ is the number of free parameters of the \mbox{$\alpha$ + core} potential. Since most levels above $8^{+}$ have uncertain spins and parities, or have not been measured, the calculation of $S.D.$~considers only the levels from $0^{+}$ to $8^{+}$ for all nuclei ($n = 5$). Most of the calculated spectra have $S.D. < 0.5$ MeV, and the $^{46}$Cr, $^{54}$Cr, and $^{68}$Se nuclei have $S.D. < 100$ keV, confirming the overall satisfactory result for the set of nuclei studied. An important feature is the existence of several nuclei with $S.D.$~close to or smaller than those found for $^{44,52}$Ti, $^{60}$Zn and $^{94}$Mo. Therefore, the calculated g.s.~bands show a good agreement with the experimental levels even for nuclei without the \mbox{$\alpha + \mathrm{DCSC}$} configuration.

\begin{table}
\begin{threeparttable}
\caption{Standard deviations ($S.D.$) associated with the calculated g.s.~bands of the nuclei studied, taking into account the levels from $0^{+}$ to $8^{+}$.}
\label{Table_errors}
\begin{ruledtabular}
\begin{tabular}{lc}
Nucleus  &	$S.D.$ (MeV)	\\[2pt] \hline
 &  \\[-6pt]
$^{44}$Ti &	0.477   \\
$^{52}$Ti & 0.668  \\
$^{46}$Cr &	0.066	\\
$^{48}$Cr & 0.375  \\
$^{54}$Cr &	0.089	\\
$^{56}$Fe &	0.304	\\
$^{58}$Fe &	0.383	\\
$^{58}$Ni &	1.108	\\
$^{60}$Ni &	0.689	\\
$^{60}$Zn &	0.155	\\
$^{82}$Zn & *  \\
$^{64}$Ge &	0.164	\\
$^{84}$Ge &	*	\\
$^{68}$Se &	0.057	\\
$^{86}$Se &	*	\\
$^{72}$Kr ($G_{\mathrm{g.s.}}=12$) &	0.721	\\
$^{72}$Kr ($G_{\mathrm{g.s.}}=14$) &	0.284	\\
$^{88}$Kr &	*	\\
$^{78}$Sr ($G_{\mathrm{g.s.}}=12$) &	1.159	\\
$^{78}$Sr ($G_{\mathrm{g.s.}}=14$) &	0.565	\\
$^{90}$Sr &	1.012	\\
$^{92}$Zr &	0.333	\\
$^{94}$Mo &	0.155	\\
\end{tabular}
\begin{tablenotes}
\item [*] Not calculated, as the experimental levels $6^{+}$ and $8^{+}$ are not measured or have undefined spin and parity.
\end{tablenotes}
\end{ruledtabular}
\end{threeparttable}
\end{table}

The $^{72}$Kr and $^{78}$Sr nuclei have been analyzed in more detail with respect to the quantum number $G_{\mathrm{g.s.}}$. As the mentioned nuclei have all experimental levels with defined spins and parities, it was possible to evaluate the most appropriate $G_{\mathrm{g.s.}}$ number for describing their g.s.~bands. In the case of $G_{\mathrm{g.s.}} = 12$, it is noted that the experimental bands of $^{72}$Kr and $^{78}$Sr are very compressed compared to the calculated bands, while in the case of $G_{\mathrm{g.s.}}$ = 14, this difference is considerably reduced (see Figs.~\ref{Fig_espec_Zn_Kr} and \ref{Fig_espec_Sr_Mo}). This feature is also noted in Table \ref{Table_errors}, where $S.D.(^{72}\mathrm{Kr})$ is reduced from 0.721 MeV to 0.284 MeV and $S.D.(^{78}\mathrm{Sr})$ is reduced from 1.159 MeV to 0.565 MeV when $G_{\mathrm{g.s.}}$ is increased from 12 to 14. This is a first indication that the number $G_{\mathrm{g.s.}}$ = 14 is the most suitable for describing the ground state bands of $^{72}$Kr and $^{78}$Sr, which suggests the $(pf)^2(sdg)^2$ configuration for valence nucleons according to the Wildermuth condition.

Shell-model calculations indicate that $^{72}$Kr has a prolate-oblate coexistence, as well as in other nuclei at or near the $N = Z$ line in the $70 \leq A \leq 80$ mass region \cite{KSW2017}. It is known that the energy sequence of shell-model orbits can undergo strong changes in deformed nuclei, including allowing the $g_{9/2}$ orbit to approach the $pf$-shell orbitals and become an intruder orbit in this shell \cite{KMS2012}. The $^{78}$Sr nucleus has $Z = 38$ and $N = 40$, i.e., two closed subshells, meaning that the excitation of the 2 valence neutrons of the $\alpha$-cluster must lead to the occupation of the $g_{9/2}$ orbit. Therefore, it is reasonable that the $(pf)^2(sdg)^2$ configuration is associated with the $^{72}$Kr and $^{78}$Sr yrast bands.

Consulting the nuclear data tables \cite{ENSDF} and more recent publications \cite{XUNDL}, it is verified that 10 of the 21 nuclei analyzed have energy levels of the g.s.~band populated in $\alpha$-transfer reactions (see Table \ref{Table_reactions}); such experimental results can be interpreted as indications of the $\alpha$-cluster structure in these nuclei. The $^{44}$Ti nucleus is the most studied by this mode, with several $\alpha$-transfer processes, where all levels of the g.s.~band have been identified. In other nuclei analyzed, the g.s.~bands are identified by $\alpha$-transfer reactions only partially, or there is no data referring to $\alpha$-transfer. The $^{90}$Sr and $^{92}$Zr nuclei have the levels from $0^{+}$ to $12^{+}$ of the g.s.~band populated through the $^{12}$C($^{86}$Kr, $2\alpha\gamma$)$^{90}$Sr and $^{88}$Sr($^{7}$Li, $2np\gamma$)$^{92}$Zr reactions, respectively. In the case of the $^{90}$Zr($^{6}$Li, $d$)$^{94}$Mo reaction, the experimental results for the $^{94}$Mo levels from $0^{+}$ to $6^{+}$ are not conclusive, due to the huge background peaks resulting from carbon and oxygen contaminants on the target, making it difficult to clearly identify the $^{94}$Mo levels. However, the $^{94}$Mo levels from $0^{+}$ to $6^{+}$ are identified through the \mbox{$^{90}$Zr($^{16}$O, $^{12}$C$\gamma$)$^{94}$Mo} reaction.

\begin{table*}
\caption{Energy levels populated in $\alpha$-transfer reactions in the ground state bands of the nuclei under study. Additionally, the energy levels populated in the $^{12}$C($^{86}$Kr,$2\alpha\gamma$)$^{90}$Sr pickup reaction (related to $^{90}$Sr) are shown. At the first mention of each experimental level, its corresponding excitation energy is indicated in MeV. Experimental data from Ref.~\cite{ENSDF}, except where indicated.}
\label{Table_reactions}
\begin{ruledtabular}
\begin{tabular}{llp{8cm}}
Nucleus     &    Reaction    &     Levels populated in the g.s.~band \\[2pt] \hline
&  &  \\[-6pt]
$^{44}$Ti   &    $^{40}$Ca($\alpha,\gamma$)       &      $0^{+}$ (g.s.), $2^{+}$ (1.083), $4^{+}$ (2.454), $6^{+}$ (4.015) \\
			&	$^{40}$Ca($^{6}$Li, $d$)      &      $0^{+}$, $2^{+}$, $4^{+}$, $6^{+}$, ($8^{+}$) (6.508) \\
			&	$^{40}$Ca(pol $^{6}$Li, $d$), ($^{6}$Li, $pn\gamma$)	 &		$0^{+}$, $2^{+}$, $4^{+}$, $6^{+}$ \\
			&	$^{40}$Ca($^{7}$Li, $t$)		&		$0^{+}$, $2^{+}$ \\
			&	$^{40}$Ca($^{12}$C, $^{8}$Be)	&		$0^{+}$, $2^{+}$ \\
			&	$^{40}$Ca($^{13}$C, $^{9}$Be), ($^{14}$N, $^{10}$B)	 &	  $0^{+}$ \\
			&	$^{40}$Ca($^{16}$O, $^{12}$C)	&		$0^{+}$, $4^{+}$, $6^{+}$, ($8^{+}$), ($10^{+}$) (7.671), ($12^{+}$) (8.040) \\
			&	$^{40}$Ca($^{20}$Ne, $^{16}$O)		&		$0^{+}$, $2^{+}$, $4^{+}$, $6^{+}$ \\
			&	$^{40}$Ca($^{32}$S, $^{28}$Si)		&		$0^{+}$ \\
			&	$^{40}$Ca($^{6}$Li, $pn\gamma$) \cite{ARB2020}    &     $0^{+}$, $2^{+}$, $4^{+}$, $6^{+}$, ($8^{+}$), ($10^{+}$), ($12^{+}$) \\
&  &  \\
$^{52}$Ti	&	$^{48}$Ca($^{6}$Li, $d$)	&		$0^{+}$ (g.s.), $2^{+}$ (1.050) \\
			&	$^{48}$Ca($^{7}$Li, $p2n\gamma$)		&		 $0^{+}$, $2^{+}$, $4^{+}$ (2.318), $6^{+}$ (3.029) \\
			&	$^{48}$Ca($^{12}$C, $^{8}$Be)	&		$0^{+}$, $2^{+}$, $4^{+}$ \\
			&	$^{48}$Ca($^{16}$O, $^{12}$C)	&		$0^{+}$, $2^{+}$ \\
&  &  \\
$^{54}$Cr	&	$^{50}$Ti($^{6}$Li, $d$)	&		$0^{+}$ (g.s.), $2^{+}$ (0.835), $4^{+}$ (1.824) \\
			&	$^{50}$Ti($^{16}$O, $^{12}$C)	&		$0^{+}$, $2^{+}$, $4^{+}$ \\
&  &  \\				
$^{56}$Fe	&	$^{52}$Cr($^{6}$Li, $d$)		&		$0^{+}$ (g.s.), $2^{+}$ (0.847), $4^{+}$ (2.085) \\
&  &  \\
$^{58}$Fe	&	$^{54}$Cr($^{6}$Li, $d$)		&		$0^{+}$ (g.s.), $2^{+}$ (0.811) \\
&  &  \\
$^{58}$Ni	&	$^{54}$Fe($^{6}$Li, $d$)	&		$0^{+}$ (g.s.), $2^{+}$ (1.454), $4^{+}$ (2.459), $6^{+}$ (5.128) \\
			&	$^{54}$Fe($^{7}$Li, $t$)		&		$0^{+}$, $2^{+}$ \\
			&	$^{54}$Fe($^{12}$C, $^{8}$Be)	&		$0^{+}$, $2^{+}$, $4^{+}$ \\
			&	$^{54}$Fe($^{16}$O, $^{12}$C)	&		$0^{+}$, $2^{+}$, $4^{+}$ \\
&  &  \\
$^{60}$Ni	&	$^{56}$Fe($^{6}$Li, $d$)	&		$0^{+}$ (g.s.), $2^{+}$ (1.333), $4^{+}$ (2.506) \\
			&	$^{56}$Fe($^{7}$Li, $2np\gamma$)		&		$0^{+}$, $2^{+}$, $4^{+}$, $6^{+}$ (4.265) \\
&  &  \\
$^{90}$Sr	&	$^{12}$C($^{86}$Kr, $2\alpha\gamma$)		&		$0^{+}$ (g.s.), $2^{+}$ (0.832), $4^{+}$ (1.656), 6$^{(+)}$ (3.495), \mbox{($8^{+}$) (5.056)}, ($10^{+}$) (5.924), ($12^{+}$) (6.795) \\
&  &  \\
$^{92}$Zr	&	$^{88}$Sr($^{7}$Li, $2np\gamma$)		&		$0^{+}$ (g.s.), $2^{+}$ (0.935), $4^{+}$ (1.495), $6^{+}$ (2.957), ($8^{+}$) (3.309), ($10^{+}$) (4.297), ($12^{+}$) (4.947)  \\
&  &  \\
$^{94}$Mo	&	$^{90}$Zr($^{16}$O, $^{12}$C$\gamma$)	&		$0^{+}$ (g.s.), $2^{+}$ (0.871), $4^{+}$ (1.574), $6^{+}$ (2.423)  \\
			&	$^{90}$Zr($^{6}$Li, $d$) \cite{YKF1998,FBL1977}		&	\emph{not conclusive}: $0^{+}$, $2^{+}$, $4^{+}$, $6^{+}$ \\
\end{tabular}
\end{ruledtabular}
\end{table*}

\begin{figure}
\includegraphics[scale=0.63]{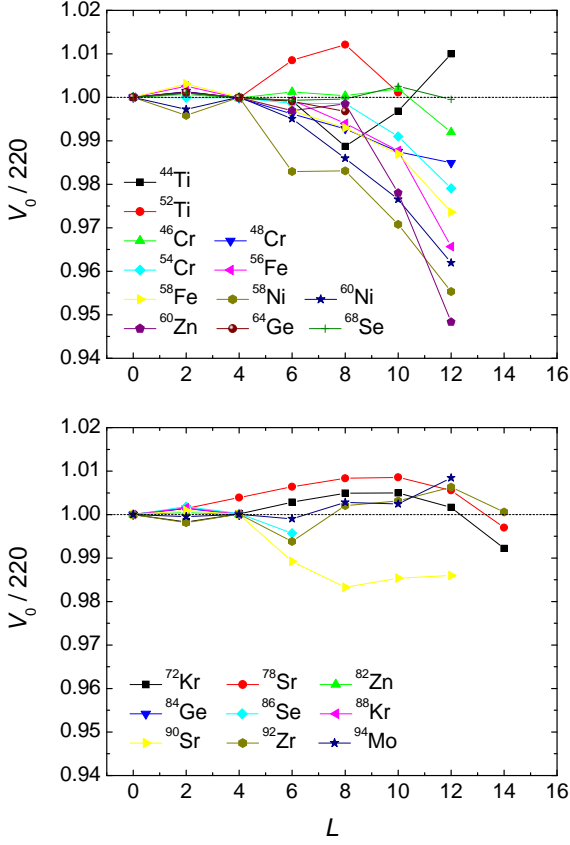}
\caption{Values of the ratio $V_{0}/220$ as a function of the quantum number $L$. The $V_0$ values have been fitted to precisely reproduce the experimental energy levels $L^{\pi}$ of the g.s.~bands. The dotted line corresponds to $V_{0}/220 = 1$. The values of $V_{0}/220$ shown for $^{72}$Kr and $^{78}$Sr correspond to $G_{\mathrm{g.s.}}=14$. (color figure online)}
\label{Fig_V0}
\end{figure}

The deviations between theoretical and experimental levels of the g.s.~bands can be analyzed in terms of a dependence of the quantum number $L$. Fig.~\ref{Fig_V0} shows graphically the values of $V_{0}/220$ as a function of $L$, where the depth parameter $V_0$ has been fitted to precisely reproduce each experimental energy level. The calculation of $V_{0}/220$ allows analyzing the relative variation of $V_0$ compared to the fixed value of 220 MeV previously applied in the description of the g.s.~bands. It is noted that the relative variation of $V_0$ is very small in general in the $0 \leq L \leq 8$ range: in most energy levels of this range, the relative variation of $V_0$ is between $-1$ \% and 1 \%. In the $10 \leq L \leq 14$ range, the variation of $V_0$ is a little greater, but still relatively small: the highest relative variations (in modulus) occur at $L = 12$ in the $^{56}$Fe, $^{58}$Ni, $^{60}$Ni, and $^{60}$Zn nuclei, between 3 \% and 5 \% approximately; such nuclei have experimental g.s.~bands with an approx.~rotational behavior, which differ more strongly from the calculated bands at the highest spin levels. Therefore, it is shown that the $\alpha$ + core potential used in this work is weakly $L$-dependent. Such dependence can not be described solely by a fitted function $V_{0}(L)$, as there is not a similar behavior among the 21 nuclei analyzed.

Table \ref{Table_radii} shows the properties calculated for the selected nuclei in the $0^{+}$ ground state: reduced $\alpha$-width ($\gamma _\alpha ^2$), dimensionless reduced $\alpha$-width ($\theta _\alpha ^2$), rms intercluster separation ($\langle R^2\rangle ^{1/2}$), the ratio of $\langle R^2\rangle ^{1/2}$ to the sum of the experimental rms charge radii of $\alpha$ and core,

\begin{equation}
\mathcal{R} = \frac{\langle R^2\rangle ^{1/2}}{\langle r^2\rangle _\alpha ^{1/2}+
\langle r^2\rangle _{\mathrm{core}}^{1/2}} \;,
\label{RR}
\end{equation}

\noindent the rms charge radius predicted by the model for the total nucleus ($\langle r^{2}\rangle_{T}^{1/2}$), the ratio of $\langle r^{2}\rangle_{T}^{1/2}$ to the corresponding experimental value (\mbox{$\langle r^{2}\rangle_{T}^{1/2}/\langle r^{2}\rangle_{T\:\mathrm{exp}}^{1/2}$}), and the ratios of $\gamma _\alpha ^2$ to the reduced $\alpha$-widths of the $\alpha + \mathrm{DCSC}$ nuclei: $^{44}$Ti, $^{60}$Zn, and $^{94}$Mo. The values of $\mathcal{R}$, $\langle r^{2}\rangle_{T}^{1/2}$ and \mbox{$\langle r^{2}\rangle_{T}^{1/2}/\langle r^{2}\rangle_{T\:\mathrm{exp}}^{1/2}$} were not calculated for some nuclei due to the lack of experimental rms radius data for the core or the total nucleus. For a better evaluation of the $G_{\mathrm{g.s.}}$ number in $^{72}$Kr, an estimate of the rms charge radius of the respective core ($^{68}$Se) is made through a proportion relation with $A^{1/3}$ and taking the experimental rms charge radius of $^{72}$Kr as reference.

\begin{turnpage}
\begin{table*}
\begin{threeparttable}
\caption{Calculated values for the reduced $\alpha $-width ($\gamma _\alpha ^2$), dimensionless reduced $\alpha $-width ($\theta _\alpha ^2$), rms intercluster separation ($\langle R^2 \rangle ^{1/2}$), the ratio $\mathcal{R}$ (see eq.~\eqref{RR}), rms charge radius predicted for the total nucleus ($\langle r^{2}\rangle_{T}^{1/2}$), the ratio of $\langle r^{2}\rangle_{T}^{1/2}$ to the corresponding experimental rms charge radius ($\langle r^{2}\rangle_{T\:\mathrm{exp}}^{1/2}$) \cite{AM2013}, and the ratios of $\gamma _\alpha ^2$ to the reduced $\alpha $-widths of $^{44}$Ti, $^{60}$Zn, and $^{94}$Mo, referring to the $0^{+}$ ground state of the nuclei studied. The channel radius used for the calculation of $\gamma _\alpha ^2$ and $\theta _\alpha ^2$ is obtained from eq.~\eqref{Eq_ch_radius} (see details in the text).}
\label{Table_radii}
\begin{ruledtabular}
\begin{tabular}{lccccccccc}
	  & $\gamma _\alpha ^2$ &	$\theta _\alpha ^2$ & $\langle R^2\rangle ^{1/2}$ &	& $\left\langle r^{2}\right\rangle_{T}^{1/2}$ &  &  &  \\[1pt]
Nucleus		& (keV)&	($10^{-3}$) &	(fm)  &	 $\mathcal{R}$ &	(fm) & 
$\left\langle r^{2}\right\rangle_{T}^{1/2}/\left\langle r^{2}\right\rangle_{T\:\mathrm{exp}}^{1/2}$ &
$\gamma _\alpha ^2\,/\,\gamma _\alpha ^2(^{44}\mathrm{Ti})$  & $\gamma _\alpha ^2\,/\,\gamma _\alpha ^2(^{60}\mathrm{Zn})$  &
$\gamma _\alpha ^2\,/\,\gamma _\alpha ^2(^{94}\mathrm{Mo})$  \\[4pt] \hline
 &  &  &  &  &  &  &  &  & \\[-6pt]
$^{44}$Ti 	  &  3.644 &   11.453 &	4.412 &	   0.8562 &	3.5858 &	0.9929	& 1.000 & 6.166 & 4.364 \\
$^{52}$Ti 	  &  1.936 &    6.178 &     4.310 &    0.8365 &	3.5756 &	---  & 0.531 &	3.276 &	2.319 \\
$^{46}$Cr	  &  1.855 &    5.968 &	4.339 &	   --- &	---	 & ---	& 0.509 &	3.139 &	2.222 \\
$^{48}$Cr	  &  0.924 &    3.107 & 4.311 &    0.8154 &	3.6891 &  ---  & 0.254 &	1.563 &	1.107 \\
$^{54}$Cr	  &  1.013 &    3.375 &	4.290 &	   0.8178 &	3.6506 &	0.9897	& 0.278 &	1.714 &	1.213 \\
$^{56}$Fe	  &  0.600 &    2.071 &	4.276 &	   0.8037 &	3.7122 &	0.9932	& 0.165 &	1.015 &	0.719 \\
$^{58}$Fe	  &  0.456 &    1.606 &	4.271 &	   0.7962 &	3.7511 &	0.9938	& 0.125 &	0.772 &	0.546 \\
$^{58}$Ni	  &  0.479 &    1.693 &	4.280 &	   0.7972 &	3.7526 &	0.9939	& 0.131 &	0.810 &	0.574 \\
$^{60}$Ni	  &  0.375 &    1.353 &	4.278 &	   0.7903 &	3.7931 &	0.9951	& 0.103 &	0.635 &	0.449 \\
$^{60}$Zn	  &  0.591 &    2.185 &	4.370 &	   --- &	--- &	---	& 0.162 &	1.000 &	0.708 \\
$^{82}$Zn	  &  0.208 &    0.850 & 4.587 &	   --- &	--- &	---	& 0.057 &	0.352 &	0.249 \\
$^{64}$Ge	  &  0.333 &    1.277 &	4.350 &	   --- &	--- &	---	& 0.091 &	0.563 &	0.399 \\
$^{84}$Ge	  &  0.232 &    0.953 &	4.607 &	   0.8083 &	4.0748 &	---	& 0.064 &	0.393 &	0.278 \\
$^{68}$Se	  &  0.194 &    0.770 &	4.335 &	   --- &	--- &	---	& 0.053 &	0.328 &	0.232 \\
$^{86}$Se	  &  0.158 &    0.674 &	4.621 &	   --- &	--- &	---	& 0.043 &	0.267 &	0.189 \\
$^{72}$Kr ($G_{\mathrm{g.s.}}=12$) &	0.119 &    0.489 &	4.328 &	   0.7513\tnote{*} &	4.1108\tnote{*} &	0.9874  & 0.033 &	0.201 &	0.143 \\
$^{72}$Kr ($G_{\mathrm{g.s.}}=14$) &  0.771 &    3.163 &	4.760 &	   0.8263\tnote{*} &	4.1358\tnote{*} &	0.9933  & 0.212 &	1.305 &	0.923 \\
$^{88}$Kr	  &  0.132 &    0.575   &	4.636 &	   --- &	--- &	---	& 0.036 &	0.223 &	0.158 \\
$^{78}$Sr ($G_{\mathrm{g.s.}}=12$) &  0.031 &    0.137 &	4.282 &	   0.7304 &	4.2036 &	0.9877	& 0.009 &	0.052 &	0.037 \\
$^{78}$Sr ($G_{\mathrm{g.s.}}=14$) &	0.233 &   1.016 &	4.712 &	   0.8038 &	4.2265 &	0.9930	& 0.064 &	0.394 &	0.279 \\
$^{90}$Sr	  &  0.135 &    0.591  &	4.646 &	   0.7930 &	4.2197 &	0.9903	& 0.037 &	0.228 &	0.162 \\
$^{92}$Zr	  &  0.154 &    0.687 &	4.682 &	   0.7936 &	4.2582 &	0.9890	& 0.042 &	0.261 &	0.184 \\
$^{94}$Mo	  &  0.835 &    3.792 &	5.104 &	   0.8586 &	4.3215 &	0.9928	& 0.229 &	1.413 &	1.000 \\
\end{tabular}
\begin{tablenotes}
\item [*] For the calculation of the indicated values, an estimate of the rms charge radius of the core is made through a
proportion relation with $A^{1/3}$ and taking as reference the experimental rms charge radius of $^{72}$Kr.
\end{tablenotes}
\end{ruledtabular}
\end{threeparttable}
\end{table*}
\end{turnpage}

The reduced $\alpha$-width is defined as \cite{AY1974,MSV1970}

\begin{equation}
\gamma _{\alpha}^2=\left( \frac{\hbar^2}{2\mu a_{c}}\right) u^2(a_c)\left[
\int_0^{a_c}|u(r)|^2dr\right] ^{-1}\;,
\label{Red_width}
\end{equation}

\noindent where $\mu $ is the reduced mass of the system, $u(r)$ is the radial wave function of the state and $a_c$ is the channel radius. The dimensionless reduced $\alpha$-width $\theta _\alpha ^2$ is defined as the ratio of $\gamma _{\alpha}^2$ to the Wigner limit,

\begin{equation}
\theta _\alpha ^2=\frac{2\mu a_{c}^2}{3\hbar ^2}\gamma _{\alpha}^2\;. 
\end{equation}

The calculation of $\gamma _{\alpha}^2$ and $\theta _\alpha ^2$ depends on the choice of the channel radius $a_c$ for each nucleus. In our previous works \cite{SM2015,SMB2019,SM2017}, $a_c$ was determined through a linear relation with \mbox{$A_{\alpha}^{1/3} + A_{\mathrm{core}}^{1/3}$}. As the present work compares isotopes with very close mass numbers, it is convenient that the channel radius varies as a function of the nuclear radius more precisely. Thus, $a_c$ is given by the equation

\begin{equation}
a_c = 1.8514 + 1.5853 \, \langle r^2\rangle _{\mathrm{core}}^{1/2} \; \mathrm{(fm)} \;.
\label{Eq_ch_radius}
\end{equation}

\noindent The parameters of eq.~\eqref{Eq_ch_radius} were fitted to minimize possible discrepancies with the $a_c$ values predicted for $^{20}$Ne, $^{44}$Ti, $^{94}$Mo, and $^{212}$Po by the formula used in Refs.~\cite{SM2015,SMB2019,SM2017}. For the cores where $\langle r^2\rangle _{\mathrm{core}}^{1/2}$ is known experimentally, $a_c$ is determined by using the tabulated values from Refs.~\cite{AM2013,LLW2021}, and for the nuclei where $\langle r^2\rangle _{\mathrm{core}}^{1/2}$ is unmeasured, the formula \cite{BAK2013}

\begin{equation}
\langle r^2\rangle^{1/2} = 0.966\textstyle{\left(1 - 0.182\frac{N - Z}{A} + \frac{1.652}{A}\right)}A^{1/3} \; \mathrm{(fm)}
\end{equation}

\noindent is applied, which describes with good accuracy the tabulated experimental rms charge radii from Ref.~\cite{AM2013}.

Analyzing the selected nuclei, it is noted that the values of $\gamma _{\alpha}^2$($0^{+}$) and $\theta _\alpha ^2$($0^{+}$) for $^{44}$Ti and $^{94}$Mo are considerably higher than the values obtained for other nuclei in their respective mass subregions (see the ratios \mbox{$\gamma _\alpha ^2\,/\,\gamma _\alpha ^2(^{44}\mathrm{Ti})$} and \mbox{$\gamma _\alpha ^2\,/\,\gamma _\alpha ^2(^{94}\mathrm{Mo})$} in Table \ref{Table_radii}), corroborating the statement that $^{44}$Ti and $^{94}$Mo are preferential nuclei for $\alpha$-clustering. This statement is reinforced by the ratio $\mathcal{R}$, since $\mathcal{R}$($^{44}$Ti) is considerably higher than the $\mathcal{R}$ values for the selected nuclei from $^{46}$Cr to $^{60}$Ni, and $\mathcal{R}$($^{94}$Mo) is considerably higher than the $\mathcal{R}$ values for the selected nuclei from $^{78}$Sr to $^{92}$Zr. The $^{60}$Zn nucleus has values of $\gamma _{\alpha}^2$($0^{+}$) and $\theta _\alpha ^2$($0^{+}$) reasonably higher than in neighboring nuclei of the set (see the ratio \mbox{$\gamma _\alpha ^2\,/\,\gamma _\alpha ^2(^{60}\mathrm{Zn})$}), indicating that its $\alpha + \mathrm{DCSC}$ condition also favors it for $\alpha$-clustering in its subregion.

It should be taken into account that $G_{\mathrm{g.s.}} = 16$ for $^{94}$Mo, while other nuclei selected in the same mass subregion have $G_{\mathrm{g.s.}} = 14$, which implies lower values of the parameter $R$ for the nuclei with smaller $G_{\mathrm{g.s.}}$ and, consequently, smaller reduced $\alpha$-widths and rms separations. However, Ref.~\cite{SM2015} indicates that $^{96}$Ru and $^{98}$Pd, which are neighbors of $^{94}$Mo and have $G_{\mathrm{g.s.}} = 16$, present $\gamma _{\alpha}^2$($0^{+}$) and $\langle R^2 \rangle ^{1/2}$($0^{+}$) values smaller than in $^{94}$Mo, corroborating the statement that $^{94}$Mo is a preferential nucleus for $\alpha$-clustering in its subregion. In addition, it should be mentioned that our recent calculation \cite{SMB2019} suggest that $^{104}$Te has an $\alpha + \mathrm{core}$ structure even more pronounced than in $^{94}$Mo, with $\gamma_{\alpha}^2(0^{+}) \approx 1$ keV.

The $\gamma _\alpha ^2\,/\,\gamma _\alpha ^2(^{44}\mathrm{Ti})$, $\gamma _\alpha ^2\,/\,\gamma _\alpha ^2(^{60}\mathrm{Zn})$ and $\gamma _\alpha ^2\,/\,\gamma _\alpha ^2(^{94}\mathrm{Mo})$ ratios are shown in Table \ref{Table_radii} for a comparison of the $\alpha$-clustering degree in the ground states of the selected nuclei in relation to $^{44}$Ti, $^{60}$Zn and $^{94}$Mo. The $\alpha + \mathrm{DCSC}$ nuclei $^{52}$Ti and $^{82}$Zn were not included as reference elements in the systematic comparison, since they have relatively lower reduced $\alpha$-widths within their mass subregions. It is convenient that the comparison of reduced $\alpha$-widths is done mainly between nuclei of the same mass subregion, as there is a trend to reduce the average magnitude of $\gamma _{\alpha}^2$ from a region of lighter mass to a region of heavier mass. In this way, it is seen there are nuclei with a considerable $\alpha$-clustering degree compared to $^{44}$Ti in its subregion, with the following $\gamma _\alpha ^2\,/\,\gamma _\alpha ^2(^{44}\mathrm{Ti})$ ratios: $^{52}$Ti (0.531), $^{46}$Cr (0.509), $^{54}$Cr (0.278), and $^{48}$Cr (0.254). The $^{48}$Cr nucleus, which was included in the comparative study exceptionally (see Sec.~\ref{Sec:Q/A}), has the smallest $\gamma _\alpha ^2\,/\,\gamma _\alpha ^2(^{44}\mathrm{Ti})$ value among the $^{46,48,54}$Cr isotopes, but it is indicated that $^{48}$Cr and $^{54}$Cr have a similar $\alpha$-clustering degree from the viewpoint of the $\alpha$ + core structure.

Compared to $^{60}$Zn, there are nuclei with a significant $\alpha$-clustering degree in its subregion, with the following $\gamma _\alpha ^2\,/\,\gamma _\alpha ^2(^{60}\mathrm{Zn})$ ratios: $^{56}$Fe (1.015), $^{58}$Fe (0.772), $^{58}$Ni (0.810), $^{60}$Ni (0.635), $^{64}$Ge (0.563), and $^{68}$Se (0.328); i.e., there are nuclei without the $\alpha + \mathrm{DCSC}$ configuration with a similar or considerable $\alpha$-clustering degree compared to $^{60}$Zn.

In the $^{94}$Mo subregion, the analyzed nuclei ($^{82}$Zn, $^{84}$Ge, $^{86}$Se, $^{88}$Kr, $^{90}$Sr, and $^{92}$Zr) have $\gamma _\alpha ^2\,/\,\gamma _\alpha ^2(^{94}\mathrm{Mo}) < 0.3$, showing a more relevant $\alpha$-clustering degree for $^{82}$Zn (0.249) and $^{84}$Ge (0.278). Although this work focuses on the $22 \leq Z \leq 42$ region, one can obtain $\gamma _\alpha ^2\,/\,\gamma _\alpha ^2(^{94}\mathrm{Mo})$ for the neighboring nuclei $^{96}$Ru and $^{98}$Pd through Ref.~\cite{SM2015} which analyzed these two nuclei in terms of the $\alpha$ + core structure with the nuclear potential of \mbox{W.S.~+ W.S.$^3$} shape; taking the values of $\gamma _{\alpha}^2(^{94}\mathrm{Mo};0^{+})$, $\gamma _{\alpha}^2(^{96}\mathrm{Ru};0^{+})$, and $\gamma _{\alpha}^2(^{98}\mathrm{Pd};0^{+})$ from Ref.~\cite{SM2015}, one obtains $\gamma _\alpha ^2\,/\,\gamma _\alpha ^2(^{94}\mathrm{Mo}) = 0.841$ and 0.702 for $^{96}$Ru and $^{98}$Pd, respectively. Therefore, the results for $^{96}$Ru and $^{98}$Pd in Ref.~\cite{SM2015} reinforce the indication in which there are nuclei around $^{94}$Mo with an expressive $\alpha$-clustering degree.

\begin{figure}
\includegraphics[scale=0.4]{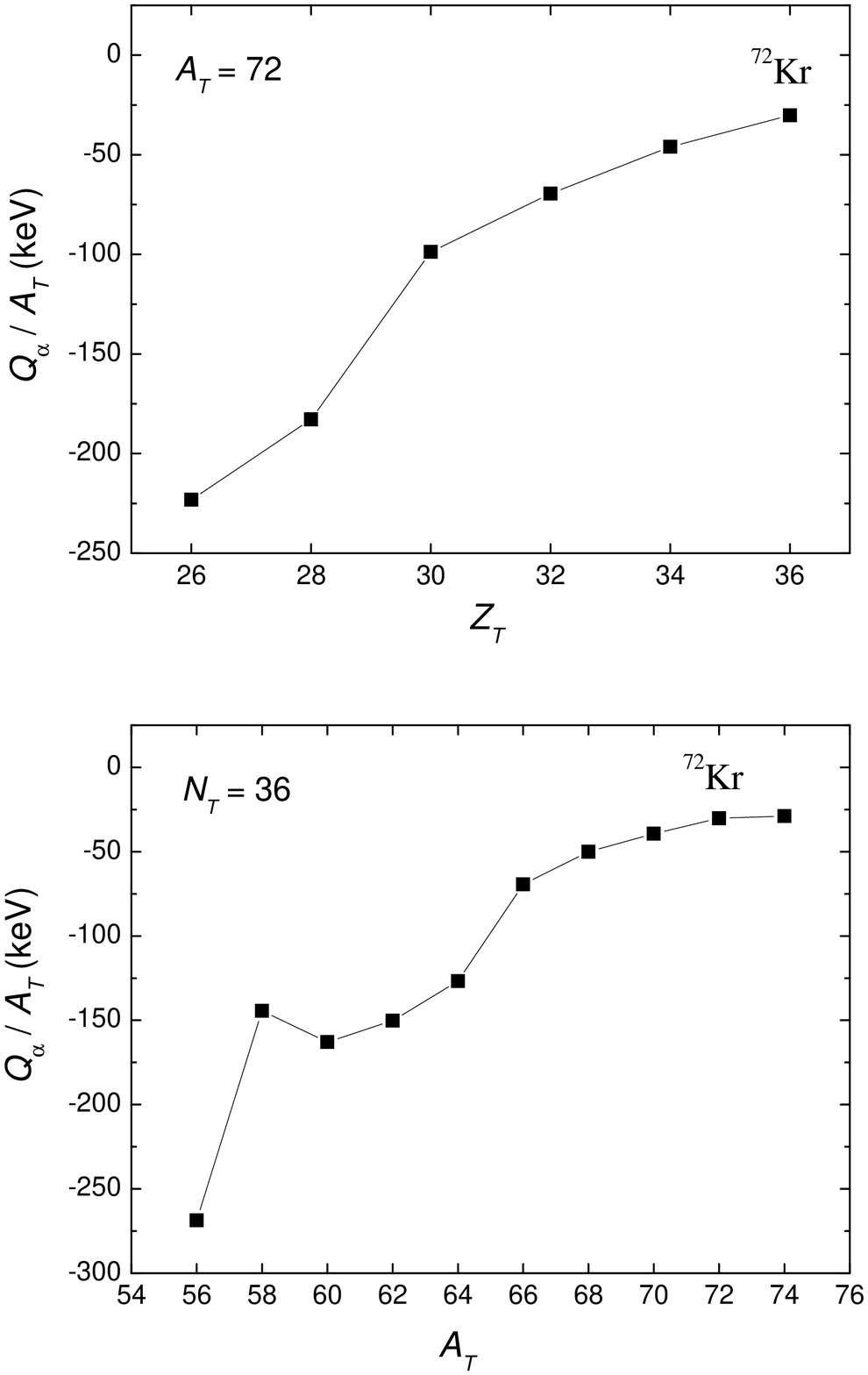}
\caption{$Q_{\alpha}/A_T$ values obtained for the $\alpha$ + core decomposition of even-even $A = 72$ isobars and $N = 36$ isotones. The $Q_{\alpha}/A_T$ values are shown as a function of the total atomic number $Z_T$ and the total mass number $A_T$.}
\label{Fig_QdivA_A72_N36}
\end{figure}

The $^{72}$Kr and $^{78}$Sr nuclei were analyzed with the band numbers $G_{\mathrm{g.s.}} = 12$ and 14, obtaining higher reduced $\alpha$-widths with $G_{\mathrm{g.s.}} = 14$. As $G_{\mathrm{g.s.}} = 14$ is more suitable for describing the g.s.~bands and rms charge radii (see discussion in the next paragraph), the analysis of the reduced $\alpha$-widths of $^{72}$Kr and $^{78}$Sr is focused on the use of $G_{\mathrm{g.s.}} = 14$. In the case of $^{78}$Sr, the ratios $\gamma _\alpha ^2\,/\,\gamma _\alpha ^2(^{60}\mathrm{Zn}) = 0.394$ and $\gamma _\alpha ^2\,/\,\gamma _\alpha ^2(^{94}\mathrm{Mo}) = 0.279$ indicate a reasonable $\alpha$-clustering degree in the $0^{+}$ ground state compared to $^{60}$Zn and $^{94}$Mo. However, in the case of $^{72}$Kr, the ratios $\gamma _\alpha ^2\,/\,\gamma _\alpha ^2(^{60}\mathrm{Zn}) = 1.305$ and $\gamma _\alpha ^2\,/\,\gamma _\alpha ^2(^{94}\mathrm{Mo}) = 0.923$ indicate a relatively high $\alpha$-clustering degree in the $0^{+}$ ground-state, even higher than $^{60}$Zn.
In addition to its favorable $Q_{\alpha}/A_T$ position among the Kr isotopes (see Fig.~\ref{Fig_QdivA}), Fig.~\ref{Fig_QdivA_A72_N36} shows that $^{72}$Kr is the nucleus with the highest $Q_{\alpha}/A_T$ value among the $A = 72$ even-even isobars and the second highest $Q_{\alpha}/A_T$ value among the $N = 36$ even-even isotones; such features point to $^{72}$Kr as one of the preferential nuclei for $\alpha$-clustering in its mass subregion. The relatively high value of $\gamma _\alpha ^2(^{72}\mathrm{Kr};0^{+})$ and the favorable $Q_{\alpha}/A_T$ position for $^{72}$Kr could be associated with a subshell closure effect due to the presence of $N_{\mathrm{core}} = 34$, or due to the proximity of $N_{\mathrm{core}} = 32$; for the time being, experimental investigations point to the existence of subshell closures at $N = 32$ and $N = 34$ in the calcium region (e.g., \cite{WBB2013,STA2013,LRA2018}). Such a suggestion needs to be reinforced by a more focused study of the nuclei neighboring $^{72}$Kr. The rough reproduction of the $B(E2)$ experimental values for $^{72}$Kr is also an aspect to be considered, as discussed further below.

The rms charge radius of the total nucleus is given by

\begin{eqnarray}
\lefteqn{ \left\langle r^2\right\rangle _{T}=\frac{Z_\alpha }{Z_\alpha +
Z_{\text{core}}}\left\langle r^2\right\rangle _\alpha + \frac{Z_{\text{core}}}
{Z_\alpha +Z_{\text{core}}}\left\langle r^2\right\rangle _{\text{core}} {} }
\nonumber\\[4pt]
 & \qquad & {} + \frac{Z_\alpha A_{\text{core}}^2+Z_{\text{core}}A_\alpha ^2}
{(Z_\alpha +Z_{\text{core}})\left( A_\alpha +A_{\text{core}}\right) ^2}
\left\langle R^2\right\rangle \;,
\label{r2_T}
\end{eqnarray}

\noindent where $\langle r^2\rangle _{T} ^{1/2}$, $\langle r^2\rangle _\alpha ^{1/2}$ and $\langle r^2\rangle _{\text{core}} ^{1/2}$ are the rms charge radii of the total system, the $\alpha$-cluster and the core, respectively. The experimental values of $\langle r^2\rangle _\alpha ^{1/2}$ and $\langle r^2\rangle _{\text{core}} ^{1/2}$ are taken from Ref.~\cite{AM2013} and $\langle R^2 \rangle ^{1/2}$ is assumed to be the calculated rms intercluster separation for the 0$^{+}$ ground state. Table \ref{Table_radii} shows the values obtained for $\langle r^{2}\rangle_{T}^{1/2}$ and \mbox{$\langle r^{2}\rangle_{T}^{1/2}/\langle r^{2}\rangle_{T\:\mathrm{exp}}^{1/2}$}. A general evaluation of the results shows there is a good agreement between theoretical and experimental rms radii, since \mbox{$\langle r^{2}\rangle_{T}^{1/2}/\langle r^{2}\rangle_{T\:\mathrm{exp}}^{1/2}$} is very close to 1 for all the nuclei in which the calculation was performed. For the $^{72}$Kr and $^{78}$Sr nuclei, $\langle r^{2}\rangle_{T}^{1/2}$ and \mbox{$\langle r^{2}\rangle_{T}^{1/2}/\langle r^{2}\rangle_{T\:\mathrm{exp}}^{1/2}$} have been calculated in the cases of $G_{\mathrm{g.s.}} = 12$ and 14. It is observed that the use of $G_{\mathrm{g.s.}} = 14$ produces a \mbox{$\langle r^{2}\rangle_{T}^{1/2}/\langle r^{2}\rangle_{T\:\mathrm{exp}}^{1/2}$} ratio closer to 1 for $^{72}$Kr and $^{78}$Sr. Therefore, in addition to the results of the spectra, the values of \mbox{$\langle r^{2}\rangle_{T}^{1/2}/\langle r^{2}\rangle_{T\:\mathrm{exp}}^{1/2}$} indicate that $G_{\mathrm{g.s.}} = 14$ is a more appropriate band number for describing the $\alpha + \mathrm{core}$ structure in $^{72}$Kr and $^{78}$Sr.

In the selected set, one notices there are nuclei with magic $N_{\mathrm{core}}$ which have reduced $\alpha$-widths smaller than their respective isotopes without such a feature. This situation occurs for the $N_{\mathrm{core}} = 50$ nuclei $^{84}$Ge, $^{86}$Se, $^{88}$Kr, and $^{90}$Sr, which have reduced $\alpha$-widths smaller than $^{64}$Ge, $^{68}$Se, $^{72}$Kr ($G_{\mathrm{g.s.}} = 14$), and $^{78}$Sr ($G_{\mathrm{g.s.}} = 14$), respectively. So, it is shown that the magic $N_{\mathrm{core}}$ condition does not necessarily determine the isotope with the most pronounced $\alpha$-cluster structure. The radial wave functions $u(r)$ tend to have smaller amplitude at the nuclear surface for lower $Q_{\alpha}$-values and in nuclei of greater mass (or greater nuclear radius), contributing to lower values of $\gamma _{\alpha}^2$ in nuclei such as $^{84}$Ge, $^{86}$Se , $^{88}$Kr and $^{90}$Sr. Additionally, $\gamma _{\alpha}^2$ decreases with the increase of the product of the $\alpha + \mathrm{core}$ reduced mass and channel radius through the factor $(\hbar^2/2\mu a_{c})$. Thus, the isotope with the highest degree of $\alpha$-clustering is defined by the combination of the previously mentioned aspects with the magic/non-magic core condition.

Another property analyzed is the $B(E2)$ transition rate between the $\alpha + \mathrm{core}$ states, given by

\begin{eqnarray}
\lefteqn{B\left(E2;G,J\rightarrow J-2\right)= {} }
                     \nonumber\\[4pt]
& & {} 
\frac{15}{8\pi}\,\beta_{2}^{2}\,\frac{J\left(J-1\right)}
{\left(2J+1\right)\left(2J-1\right)}\left\langle
 r_{J,J-2}^{2}\right\rangle ^{2}\;,
\label{BE2_a}
\end{eqnarray}

\noindent where

\begin{equation}
\left\langle r_{J,J-2}^{2}\right\rangle =
\int_{0}^{\infty}r^{2}\, u_{G,J}(r)\, u_{G,J-2}(r)dr \;, \label{BE2_b}
\end{equation}

\noindent $\beta_{2}$ is the recoil factor, given by

\begin{equation}
\beta_{2}=\frac{Z_{\alpha}A_{\mathrm{core}}^{2}+Z_{\mathrm{core}}A_{\alpha}^{2}}
{\left(A_{\alpha}+A_{\mathrm{core}}\right)^{2}}\;,
\label{BE2_c}
\end{equation}

\noindent $u_{G,J}(r)$ and $u_{G,J-2}(r)$ are the radial wave functions of the
initial $|G,J\rangle$ state and final $|G,J-2\rangle$ state, respectively.

The calculated $B(E2)$ values for the $2^{+} \rightarrow 0^{+}$ and $4^{+} \rightarrow 2^{+}$ transitions, \emph{without} the use of effective charges, are shown in Table \ref{Table_B(E2)} in comparison with experimental data. For some nuclei, $B(E2)$ is calculated only for the $2^{+} \rightarrow 0^{+}$ transition when there is not experimental value available for $4^{+} \rightarrow 2^{+}$. When calculating the functions $u_{G,J}(r)$, the depth $V_{0}$ was slightly modified to reproduce precisely the corresponding experimental energy levels (see $V_{0}$ values in Table \ref{Table_B(E2)}). For comparison, the respective $B(E2)$ values obtained in previous shell-model calculations are shown in Table \ref{Table_B(E2)} (refs.~indicated in the table). Comparing the calculated and experimental values, it is observed that most of the calculated $B(E2)$ values reproduce the order of magnitude of the respective experimental data. In the cases of $^{52}$Ti, $^{58}$Ni, $^{90}$Sr, and $^{92}$Zr, the calculated $B(E2)$ rates are relatively close to the respective experimental values. In general, the results obtained in the shell-model calculations are closer to the experimental $B(E2)$ data; however, it should be taken into account that the shell-model results presented in Table \ref{Table_B(E2)} were obtained with the use of high effective charges in some cases: for protons, the effective charge $e_{\pi}$ ranges from $1.5 \, e$  to $2.32 \, e$, and for neutrons, the effective charge $e_{\nu}$ ranges from $0.5 \, e$ to $1.73 \, e$. Therefore, considering that effective charges were not used in the calculations of this work, the $B(E2)$ results obtained by the $\alpha$ + core model are at least satisfactory.

\begin{table*}
\begin{threeparttable}
\caption{$B(E2)$ rates calculated for the $2^{+}_1 \rightarrow 0^{+}_1$ and $4^{+}_1 \rightarrow 2^{+}_1$ transitions in the g.s.~bands of the nuclei studied, \emph{without} the use of effective charges. The table shows the depths $V_{0}$ fitted to reproduce precisely the experimental energy levels. Experimental data are from Ref.~\cite{ENSDF}, except where indicated. The corresponding $B(E2)$ values obtained from shell-model calculations are shown with the respective references.}
\label{Table_B(E2)}
\begin{ruledtabular}
\begin{tabular}{cccccc}
  &	 &	$V_{0}$ &	\multicolumn{3}{c}{$B(E2;J\rightarrow J-2)$ (W.u.)} \\[2pt] \cline{4-6}
 &  &  &  &  &  \\[-7pt]
 Nucleus &	$J^\pi$ &	(MeV) &	This work & Expt. & Shell-model\tnote{{\it a}} \\[2pt] \hline
 &  &  &  &  &  \\[-6pt]
$^{44}$Ti &	$0^{+}$ &	220.00 &     &     &     \\
  &	$2^{+}$ &	220.22 &	10.810 & 22.2($+22$ $-18$) \cite{ABS2017} &	24.5 \cite{ARB2020} \tnote{{\it c}} \\
  &	$4^{+}$ &	219.98 &	14.781 & 30($+4$ $-3$) \cite{ARB2020}  &	 35.4 \cite{ARB2020} \tnote{{\it c}} \\
  &     &     &     &     &     \\
$^{52}$Ti &	$0^{+}$ &	220.00 &     &     &     \\
  &	$2^{+}$ &	220.27 &	7.909 &	7.5($+4$ $-3$) \cite{GFB2019} & 8.7 \cite{NSO1994} \\
  &	$4^{+}$ &	219.97 &	10.774 & 9.4($+14$ $-11$) \cite{GFB2019} & 11.6 \cite{NSO1994} \\
  &     &     &     &     &     \\
$^{46}$Cr &	$0^{+}$ &	220.00 &     &     &     \\
  &	$2^{+}$ &	220.14 &	9.657 &	19(4) &	18.7 \cite{M2008_RJP} \tnote{{\it d}} \\
  &     &     &     &     &     \\
$^{48}$Cr &	$0^{+}$ &	220.00 &     &     &     \\
  &	$2^{+}$ &	220.28 &	8.858 &	 27($+2$ $-1$) \cite{ABS2017} & 22.0 \cite{CZP1994} \\
  &	$4^{+}$ &	219.97 &	11.950 &	27(3) & 30.1 \cite{CZP1994} \\
  &     &     &     &     &     \\
$^{54}$Cr &	$0^{+}$ &	220.00 &     &     &     \\
 &	$2^{+}$ &	219.98 &	7.456 &	14.4(6) &	14.8 \cite{NSO1994}  \\
 &	$4^{+}$ &	219.98 &	10.049 &	26(9) &	19.6 \cite{NSO1994}  \\
  &     &     &     &     &     \\
$^{56}$Fe &	$0^{+}$ &	220.00 &     &     &     \\
  &	$2^{+}$ &	220.57 &	7.106 &	16.8(7) &	12.9 \cite{NSO1994}  \\
  &	$4^{+}$ &	219.96 &	9.648 &	24(5) &	18.0 \cite{NSO1994}  \\
  &     &     &     &     &     \\
$^{58}$Fe &	$0^{+}$ &	220.00 &     &     &     \\
 &	$2^{+}$ &	220.68 &	6.748 &	18.5(6) &	16.9 \cite{MR1979}  \\
 &	$4^{+}$ &	220.01 &	9.168 &	47(7) &	16.7 \cite{MR1979}  \\
  &     &     &     &     &     \\
$^{58}$Ni &	$0^{+}$ &	220.00 &     &     &     \\
 &	$2^{+}$ &	219.09 &	6.946 &	9.4(6) \cite{ABS2014} \tnote{{\it b}} &	3.9 \cite{NSO1994}  \\
 &	$4^{+}$ &	220.00 &	9.496 &	3.7($+8$ $-4$) \cite{LIB2016} &	2.9 \cite{NSO1994}  \\
  &     &     &     &     &     \\
$^{60}$Ni &	$0^{+}$ &	220.00 &     &     &     \\
 &	$2^{+}$ &	219.39 &	6.615 &	13.0(6) \cite{ABS2014} \tnote{{\it b}} &	12.9 \cite{PSP1981}  \\
 &	$4^{+}$ &	220.01 &	9.072 &	(5.5(17)) &	5.7 \cite{PSP1981}  \\
  &     &     &     &     &     \\
$^{60}$Zn &	$0^{+}$ &	220.01 &     &     &     \\
 &	$2^{+}$ &	220.22 &	7.252 &	--- &    \\
  &     &     &     &     &     \\
$^{82}$Zn &	$0^{+}$ &	220.00 &     &     &     \\
 &  $2^{+}$ &	220.10 &	5.831 &	--- &    \\
  &     &     &     &     &     \\
$^{64}$Ge  &	$0^{+}$ &	220.00 &     &     &     \\
 &	$2^{+}$ &	220.26 &	6.574 &	27(4) \cite{SDD2007} &	26.7 \cite{SDD2007}   \\
  &     &     &     &     &     \\
$^{84}$Ge &	$0^{+}$ &	220.00 &     &     &     \\
 &	$2^{+}$ &	220.31 &	5.790 &	28($+70$ $-10$) \cite{DVM2018} &	 17.0 \cite{SRK2013}  \\
 &	$4^{+}$ &	220.00 &	7.926 &	11($+18$ $-2$) \cite{DVM2018} &	16.6 \cite{SRK2013}  \\
\end{tabular}
\begin{tablenotes}
\item [{\it a}] All shell-model calculations mentioned apply effective charges to protons ($e_{\pi}$) and neutrons ($e_{\nu}$). For the values shown, $e_{\pi}$ ranges from $1.5 \, e$  to $2.32 \, e$, and $e_{\nu}$ ranges from $0.5 \, e$ to $1.73 \, e$ (see specific values in references).
\item [{\it b}] Obtained from the experimental value of $B(E2; 0_{1}^{+} \rightarrow 2_{1}^{+})$.
\item [{\it c}] Results obtained with the ZBM2M interaction.
\item [{\it d}] Results obtained with the GXPF1 interaction.
\end{tablenotes}
\end{ruledtabular}
\end{threeparttable}
\end{table*}

\begin{table*}[t!]
\begin{center}
{\small TABLE \ref{Table_B(E2)}. (continued)}
\end{center}
\vspace{-2mm}
\begin{ruledtabular}
\begin{tabular}{cccccc}
  &	 &	$V_{0}$ &	\multicolumn{3}{c}{$B(E2;J\rightarrow J-2)$ (W.u.)} \\[2pt] \cline{4-6}
 &  &  &  &  &  \\[-7pt]
 Nucleus &	$J^\pi$ &	(MeV) &	This work & Expt. & Shell-model$^{a}$ \\[2pt] \hline
 &  &  &  &  &  \\[-6pt]
$^{68}$Se &	$0^{+}$ &	220.00 &     &     &     \\
 &	$2^{+}$ &	220.21 &	6.022 &	24(4) \cite{NWI2014} &	30.5 \cite{KMS2009}  \\
  &     &     &     &     &     \\
$^{86}$Se &	$0^{+}$ &	220.00 &     &     &     \\
 &	$2^{+}$ &	220.41 &	5.718 &	20($+5$ $-2$) \cite{DVM2018} &	19.3 \cite{SRK2013}  \\
 &	$4^{+}$ &	220.05 &	7.842 &	$\geq$ 5.7 & 19.5 \cite{SRK2013}  \\
  &     &     &     &     &     \\
$^{72}$Kr ($G_{\mathrm{g.s.}}=12$) &	$0^{+}$ &	220.02 &     &     &     \\
 &	$2^{+}$ &	219.51 &	5.545 &	45(8) \cite{ILM2014} &	1.6 \cite{KSW2017}, 19.1 \cite{HKM2007}  \\
 &	$4^{+}$ &	220.49 &	7.363 &	153(31) \cite{ILM2014} & 128.5 \cite{KSW2017}  \\
  &     &     &     &     &     \\
$^{72}$Kr ($G_{\mathrm{g.s.}}=14$) &	$0^{+}$ &	220.00 &     &     &     \\
 &	$2^{+}$ &	219.62 &	8.161 &	45(8) \cite{ILM2014} &	1.6 \cite{KSW2017}, 19.1 \cite{HKM2007}  \\
 &	$4^{+}$ &   220.03 &	11.060 & 153(31) \cite{ILM2014} &	128.5 \cite{KSW2017}  \\
  &     &     &     &     &     \\
$^{88}$Kr &	$0^{+}$ &	220.00 &     &     &     \\
 &	$2^{+}$ &	220.33 &	5.651 &	12($+10$ $-4$) \cite{DVM2018} &	14.1 \cite{SRK2013}  \\
  &     &     &     &     &     \\
$^{78}$Sr ($G_{\mathrm{g.s.}}=12$) &	$0^{+}$ &	220.04 &     &     &     \\
 &	$2^{+}$ &	220.66 &	4.761 &	93(5) \cite{LBW2020} &	126 \cite{TS1994}  \\
 &	$4^{+}$ &	221.74 &	6.300 &	169(17) &	192 \cite{TS1994}  \\
  &     &     &     &     &     \\
$^{78}$Sr ($G_{\mathrm{g.s.}}=14$) &	$0^{+}$ &	220.02 &     &     &     \\
 &	$2^{+}$ &	220.31 &	7.015 &	93(5) \cite{LBW2020} &	126 \cite{TS1994}  \\
 &	$4^{+}$ &	220.86 &	9.450 &	169(17) &	192 \cite{TS1994} \\
  &     &     &     &     &     \\
$^{90}$Sr &	$0^{+}$ &	220.00 &     &     &     \\
 &	$2^{+}$ &	220.20 &	5.565 &	8.5($+33$ $-19$) &	9.7 \cite{SSR2001}  \\
 &	$4^{+}$ &	220.01 &	7.634 &	5.2($+11$ $-7$) &	5.5 \cite{SSR2001}  \\
  &     &     &     &     &     \\
$^{92}$Zr &	$0^{+}$ &	220.00 &     &     &     \\
 &	$2^{+}$ &	219.59 &	5.619 &	6.18(23) \cite{OBC2013} &	6.0 \cite{WBB2002}  \\
 &	$4^{+}$ &	220.02 &	7.662 &	4.05(12) &	4.4 \cite{WBB2002}  \\
  &     &     &     &     &     \\
$^{94}$Mo &	$0^{+}$ &	220.00 &     &     &     \\
 &	$2^{+}$ &	219.90 &	7.742 &	16.0(4) &	16.5 \cite{LPF2000}  \\
 &	$4^{+}$ &	220.01 &	10.742 &	26(4) &	17.5 \cite{LPF2000}  \\
\end{tabular}
\end{ruledtabular}
\end{table*}

The $B(E2)$ rates for the $^{72}$Kr and $^{78}$Sr nuclei were calculated in the cases $G_{\mathrm{g.s.}}$ = 12 and 14 to make a comparison of the results. The experimental $B(E2)$ rates for $^{72}$Kr and $^{78}$Sr are significantly higher than in the other selected nuclei. Again, the number $G_{\mathrm{g.s.}} = 14$ is shown to be more suitable for the two nuclei, since this band number produces reasonably higher $B(E2)$ rates; however, the increase in $B(E2)$ produced by $G_{\mathrm{g.s.}} = 14$ is still insufficient for a satisfactory reproduction of the experimental data. If an effective charge $\delta e$ is applied\footnote{The effective charge is defined so that $e_{\nu} = \delta e$ and $e_{\pi} = \delta e + e$.} for the precise reproduction of the experimental $B(E2; 2^{+} \rightarrow 0^{+})$ rate, its value would be $0.674 \, e$ for $^{72}$Kr ($G_{\mathrm{g.s.}} = 14$) and $1.321 \, e$ for $^{78}$Sr ($G_{\mathrm{g.s.}} = 14$), meaning a high effective charge for $^{78}$Sr. The strong increase in the experimental $B(E2)$ rates and the decrease in the experimental energy spacing $2^{+}_1 \rightarrow 0^{+}_1$ indicate the increase of collectivity in the transitions in these two nuclei, especially in $^{78}$Sr \cite{RSB2013}. Therefore, the strong difference between calculated and experimental $B(E2)$ values suggests that the single contribution of the $\alpha + \mathrm{core}$ structure in the $2^{+}_1 \rightarrow 0^{+}_1$ and $4^{+}_1 \rightarrow 2^{+}_1$ transitions is not sufficient to describe the collective effects manifested in $^{72}$Kr and $^{78}$Sr.

\begin{figure}[h!]
\includegraphics[scale=0.6]{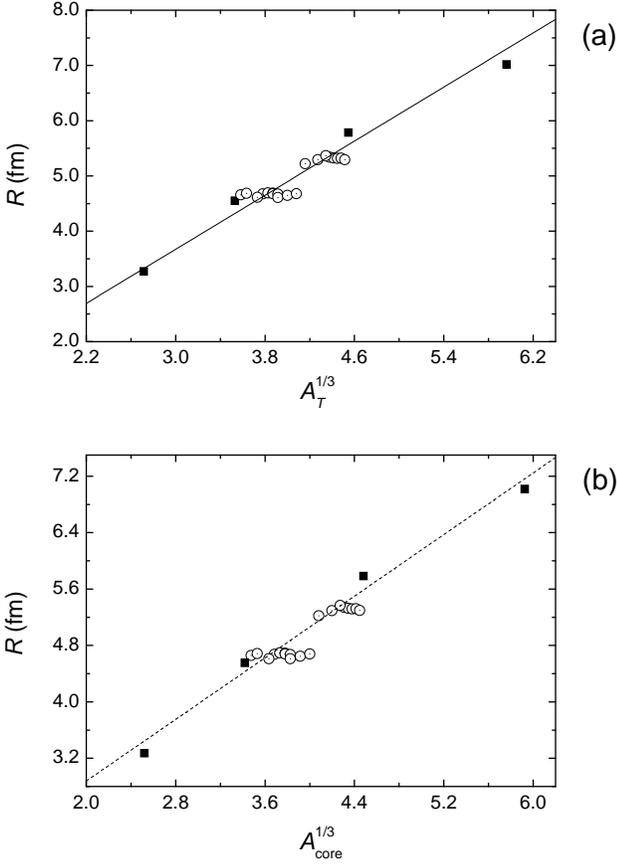}
\caption{Values of the parameter $R$ as a function of $A_{T}^{1/3}$ {\bf (a)} and $A_{\mathrm{core}}^{1/3}$ {\bf (b)}, where $A_T$ and $A_{\mathrm{core}}$ are the mass numbers of the total nucleus and the core, respectively. The full squares correspond to the set \{$^{20}$Ne, $^{44}$Ti, $^{94}$Mo, $^{212}$Po\}; the open circles correspond to the set from $^{46}$Cr to $^{92}$Zr analyzed in this work. The $R$ values shown in the graphs for $^{72}$Kr and $^{78}$Sr correspond to $G_{\mathrm{g.s.}}=14$. The full and short dashed lines correspond to the linear fits described by eqs.~\eqref{Eq_R_AT} and \eqref{Eq_R_Acore}, respectively.}
\label{Fig_R_fits}
\end{figure}

An analysis of the parameter $R$ is made in relation to $A_T^{1/3}$ and $A_{\mathrm{core}}^{1/3}$. Fig.~\ref{Fig_R_fits} shows the values of $R$ for the set of nuclei analyzed in this work along with $^{20}$Ne and $^{212}$Po. The $R$ values for $^{20}$Ne and $^{212}$Po were obtained from our previous work \cite{SMB2019} on the same $\alpha + \mathrm{core}$ potential applied to nuclei with $\alpha$-clustering above double shell closures. The full squares correspond to the set \{$^{20}$Ne, $^{44}$Ti, $^{94}$Mo, $^{212}$Po\} and the open circles correspond to the nuclei from $^{46}$Cr to $^{92}$Zr selected in this work. Due to the more favorable results for $^{72}$Kr and $^{78}$Sr by using $G_{\mathrm{g.s.}} = 14$, only the $R$ values related to $G_{\mathrm{g.s.}} = 14$ are shown graphically for these two nuclei. Fig.~\ref{Fig_R_fits} shows the linear trend of the points presented as a function of $A_T^{1/3}$ and $A_{\mathrm{core}}^{1/3}$. An analysis option would be to make a linear fit involving all nuclei, including the set \{$^{20}$Ne, $^{44}$Ti, $^{94}$Mo, $^{212}$Po\}; however, such an option would result in a strongly biased fit for the $22 \leq Z \leq 42$ region, and weakly influenced by the mass regions of $^{20}$Ne and $^{212}$Po. Therefore, the analysis of the $R$ values was made using the linear fits presented in Ref.~\cite{SMB2019} for the set \{$^{20}$Ne, $^{44}$Ti, $^{94}$Mo, $^{212}$Po\}:

\begin{equation}
R = 1.224 \, A_{T}^{1/3} \; \mathrm{(fm)}
\label{Eq_R_AT}
\end{equation}

\noindent and

\begin{equation}
R = 0.694 + 1.092 \, A_{\mathrm{core}}^{1/3} \; \mathrm{(fm).}
\label{Eq_R_Acore}
\end{equation}

\noindent Through eqs.~\eqref{Eq_R_AT} and \eqref{Eq_R_Acore}, it is possible to verify whether the set from $^{46}$Cr to $^{92}$Zr behaves naturally according to the linear trend previously observed in different mass regions.

To more accurately assess the linear behavior of $R$ as a function of $A_T^{1/3}$ and $A_{\mathrm{core}}^{1/3}$, Table \ref{Table_R_fits} shows the values obtained for the Pearson's correlation coefficient ($r_P$) for two sets: the first, containing the nuclei from $^{46}$Cr to $^{92}$Zr  selected in this work, and the second, containing $^{20}$Ne, $^{44}$Ti, $^{94}$Mo, and $^{212}$Po. The set from $^{46}$Cr to $^{92}$Zr has $r_P \approx 0.91$ in relation to $A_T^{1/3}$ and $A_{\mathrm{core}}^{1/3}$, and the set \{$^{20}$Ne, $^{44}$Ti, $^{94}$Mo, $^{212}$Po\} has $r_P \approx 0.99$ in relation to the same variables, showing that the linear trend is strong both locally and in nuclei of different mass regions. In the $22 \leq Z \leq 42$ region, there are two groups of nuclei distanced by $\approx 0.6$ fm on the $R$ scale, corresponding to the quantum numbers $G_{\mathrm{g.s.}} = 12$ and 14, while $^{94}$Mo is $\approx 0.5$ fm above the $G_{\mathrm{g.s.}}$ = 14 group since it corresponds to $G_{\mathrm{g.s.}}$ = 16. Therefore, there is a more abrupt increase in $R$ at the transition from a $G_{\mathrm{g.s.}}$ number to another band number just above, which do not significantly affect the general linear behavior of $R$.

The standard deviation of the $R$ values related to eqs.~\eqref{Eq_R_AT} and \eqref{Eq_R_Acore} are calculated for the set \{$^{20}$Ne, $^{44}$Ti, $^{94}$Mo, $^{212}$Po\} and the set of nuclei selected from $^{46}$Cr to $^{92}$Zr (Table \ref{Table_R_fits}). In the case of the set \{$^{20}$Ne, $^{44}$Ti, $^{94}$Mo, $^{212}$Po\}, $S.D. = 0.246$ fm and 0.229 fm in relation to eqs.~\eqref{Eq_R_AT} and \eqref{Eq_R_Acore}, respectively. In the case of the set from $^{46}$Cr to $^{92}$Zr, $S.D. = 0.162$ fm and 0.188 fm in relation to eqs.~\eqref{Eq_R_AT} and \eqref{Eq_R_Acore}, respectively, representing a small relative error for the parameter $R$ which varies from $\approx 4.6$ fm to $\approx 5.4$ fm in this set. Therefore, the linear relation of $R$ with $A_T ^{1/3}$ and $A_{\mathrm{core}}^{1/3}$ is verified both locally ($22 \leq Z \leq 42$) and in nuclei of different mass regions.

\begin{table*}
\caption{Pearson's correlation coefficient ($r_P$) and standard deviation ($S.D.$) for parameter $R$ in relation to the variables $A_T ^{1/3}$ and $A_{\mathrm{core}}^{1/3}$ and the linear fits described in eqs.~\eqref{Eq_R_AT} and \eqref{Eq_R_Acore}. The properties are calculated for the set \{$^{20}$Ne, $^{44}$Ti, $^{94}$Mo, $^{212}$Po\} and the set from $^{46}$Cr to $^{92}$Zr analyzed in this work. The calculations take the $R$ values for $^{72}$Kr and $^{78}$Sr corresponding to $G_{\mathrm{g.s.}}=14$.}
\label{Table_R_fits}
\begin{ruledtabular}
\begin{tabular}{lcccc}
Set of nuclei & Related variable &  Fitted function & $r_P$ & $S.D.$ (fm) \\[2pt] \hline
&  &  &  &  \\[-6pt]
$^{46}$Cr to $^{92}$Zr & $A_{\mathrm{core}}^{1/3}$ & eq.~\eqref{Eq_R_Acore} &  0.9054  &  0.188 \\[2pt]
$^{46}$Cr to $^{92}$Zr & $A_T ^{1/3}$ & eq.~\eqref{Eq_R_AT} &  0.9062  &  0.162 \\[2pt]
$^{20}$Ne, $^{44}$Ti, $^{94}$Mo, $^{212}$Po & $A_{\mathrm{core}}^{1/3}$ & eq.~\eqref{Eq_R_Acore} & 0.9932 & 0.229 \\[2pt]
$^{20}$Ne, $^{44}$Ti, $^{94}$Mo, $^{212}$Po & $A_T ^{1/3}$ & eq.~\eqref{Eq_R_AT} & 0.9912 & 0.246 \\
\end{tabular}
\end{ruledtabular}
\end{table*}

\begin{figure}
\includegraphics[scale=0.34]{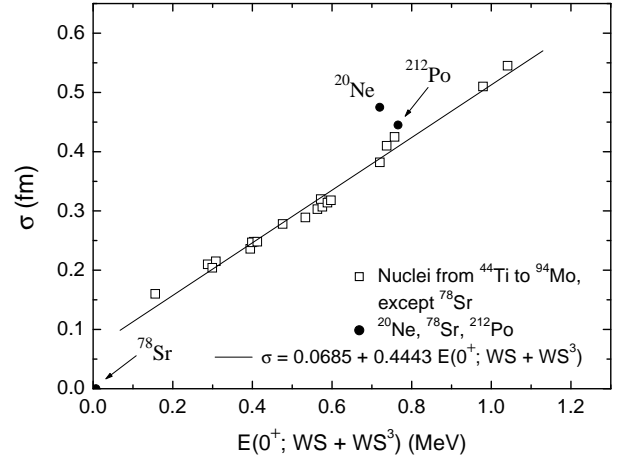}
\caption{Values of the parameter $\sigma$ as a function of the theoretical energy of the $0^{+}$ state produced by the \mbox{W.S.~+ W.S.$^3$} potential. The $\sigma$ values presented for $^{72}$Kr and $^{78}$Sr correspond to $G_{\mathrm{g.s.}}=14$. The full line shows the linear fit for the nuclei selected from $^{44}$Ti to $^{94}$Mo, except $^{78}$Sr (open squares - see details in the text). The $\sigma$ values for $^{20}$Ne, $^{78}$Sr and $^{212}$Po are indicated by full circles.}
\label{Fig_Sigma}
\end{figure}

This paragraph discusses the role of the free parameter $\sigma$ in the \mbox{(1 + Gaussian)$\times$(W.S.~+ W.S.$^3$)} potential. $\sigma$ has the function of correctly adjusting the $0^{+}$ ground state at the $\alpha + \mathrm{core}$ separation energy ($Q_{\alpha}$). It is known that the \mbox{W.S.~+ W.S.$^3$} nuclear potential provides a satisfactory description in general of the g.s.~bands in several nuclei \cite{BMP95,SM2015}; however, the experimental $0^{+} \rightarrow 2^{+}$ energy spacing is somewhat roughly described, as it is generally incompatible with the rotational behavior that the \mbox{W.S.~+ W.S.$^3$} potential produces at the first levels of the band. Fig.~\ref{Fig_Sigma} shows the relation of $\sigma$ with the $0^{+}$ state energy produced by the \mbox{W.S.~+ W.S.$^3$} nuclear potential, using the $R$ values from Table \ref{Table_parameters} and the same fixed parameters $a$, $b$ and $V_0$. In general, the energy \mbox{$E$($0^{+}$; W.S.~+ W.S.$^3$)} is above that predicted by the $\alpha + \mathrm{core}$ separation (from a few tenths of MeV to $\approx 1$ MeV) since, as in Refs.~\cite{BMP95,SM2015}, priority is given to more accurate reproduction of the experimental $4^{+}$ level. The $\sigma$ values of the selected nuclei from $^{44}$Ti to $^{94}$Mo (except $^{78}$Sr) are used for a linear fit as a function of \mbox{$E$($0^{+}$; W.S.~+ W.S.$^3$)}. The fit shown in Fig.~\ref{Fig_Sigma} and the corresponding correlation coefficient $r_P = 0.992$ demonstrate the strong linear trend of $\sigma$ as a function of \mbox{$E$($0^{+}$; W.S.~+ W.S.$^3$)} in the $22 \leq Z \leq 42$ region. The $\sigma$ value for $^{78}$Sr is not included in the fit because it was established beforehand as $\sigma = 0$; in this case, the corresponding experimental spacing $0^{+} \rightarrow 2^{+}$ is compatible with a rotational spectrum (see Fig.~\ref{Fig_espec_Sr_Mo}), being suitable to cancel the effect of the \mbox{(1 + Gaussian)} factor for $^{78}$Sr. 

The $\sigma$ values for $^{20}$Ne and $^{212}$Po are shown as full circles in Fig.~\ref{Fig_Sigma}. Note that these two points are slightly distant from the fitted line, mainly $^{20}$Ne, indicating that the linear fit of $\sigma$ as a function of \mbox{$E$($0^{+}$; W.S.~+ W.S.$^3$)} should be reevaluated in other mass regions. Despite this, the analysis of the parameter $\sigma$ clarifies that it becomes greater when the theoretical level \mbox{$E$($0^{+}$; W.S.~+ W.S.$^3$)} deviates more strongly from the experimental $0^{+}$ bandhead ($E_x = 0$).

\section{Discussion on the \mbox{$\alpha$ + core} negative parity bands}

The \mbox{(1 + Gaussian)$\times$(W.S.~+ W.S.$^3$)} potential supports excited bands, such as the first negative parity band ($G = G_{\mathrm{g.s.}} + 1$) and the second positive parity band ($G = G_{\mathrm{g.s.}} + 2$). The calculation of the properties of the excited bands involves greater complications compared to the g.s.~bands, as many experimental levels at higher energies have undefined spins and parities, or often there are not experimental energy levels corresponding to the calculated levels. The present work focused on the analysis of the g.s.~bands; however, the results presented here serve as a start point for a next study on the excited bands. At this stage, a discussion on the energy location of the negative parity bands is presented for some nuclei of this study in comparison with previous calculations. In the predictions for the negative parity bands below, we use the fixed values of $V_0$, $a$, $b$, and $\lambda$ (except where other parameter values are explicitly mentioned), and the values of $R$ and $\sigma$ from Table \ref{Table_parameters}.

The existence of a $G = 13$ negative parity band in $^{44}$Ti has already been discussed in previous theoretical works (examples in Refs.~\cite{MRO88,O1988,MPH1989,MOR1998,IMP2019,KSB2017}). The $\alpha + ^{40}$Ca potential used in this work produces a $1^{-}$ bandhead at $E_x = 7.366$ MeV. This value is similar to the previous calculations in Refs.~\cite{MRO88,O1988,MPH1989,MOR1998,IMP2019,KSB2017}, where the $1^{-}$ state is in the range of $6 \; \mathrm{MeV} \lessapprox E_x \lessapprox 8 \; \mathrm{MeV}$. The experimental investigation of $^{44}$Ti through the \mbox{$^{40}$Ca($^{6}$Li, {\it d})$^{44}$Ti} \cite{YOF1990,GJZ1993,YII1996,YKF1998} and \mbox{$^{40}$Ca($^{7}$Li, $t \alpha$)$^{40}$Ca} \cite{FTO2009} reactions identified experimental levels from $J^{\pi} = 1^{-}$ to $7^{-}$ which are candidates for members of the $G = 13$ band, where the experimental level $J^{\pi} = 1^{-}$ lies at $E_x = 6.22$ MeV.

In the case of $^{52}$Ti, the $\alpha + ^{48}$Ca potential of this work produces $E_{x}(1^{-}) = 7.842$ MeV for the $G = 13$ band. This value is above that predicted by Ohkubo \cite{O2020} ($E_{x}(1^{-}) = 6.90$ MeV ), which uses a Woods-Saxon squared local potential for the $\alpha$ + core nuclear interaction. The $\alpha$ + core nuclear potential of Ref.~\cite{O2020} was obtained from the real part of an optical potential applied in the analysis of the $\alpha$-particle scattering from $^{48}$Ca, using a depth parameter fitted properly to reproduce the binding energy of the $^{52}$Ti ground state. Additional experimental data are needed for the characterization of the $G = 13$ band of $^{52}$Ti.

The $\alpha$ + core structure in the $^{46}$Cr and $^{54}$Cr isotopes was analyzed by the present authors \cite{SM2017} and Mohr \cite{M2017}, with emphasis on the properties of the $G = 12$ g.s.~band. Using the $\alpha$ + core potential of the present work, the $1^{-}$ bandhead for the $G = 13$ band is obtained at $E_x = 7.806$ MeV for $^{46}$Cr and $E_x = 7.902$ MeV for $^{54}$Cr. The $^{46}$Cr experimental spectrum presents the levels ($3^{-}$)($E_x = 3.196$ MeV), ($5^{-}$)($E_x = 3.987$ MeV), and ($7^{-}$)($E_x = 5.346$ MeV), which can be related to a negative parity band; however, the theoretical prediction $E_x(1^{-}) = 7.806$ MeV is incompatible with the experimental levels mentioned. In the case of $^{54}$Cr, there are not experimental energy levels which can characterize a negative parity band to be compared with theoretical predictions.

In the case of $^{48}$Cr, the $\alpha + ^{44}$Ti potential of this work produces $E_x(1^{-}) = 7.896$ MeV for the $G = 13$ band. In the calculation of Descouvemont on $^{48}$Cr \cite{D2002}, which was analyzed in terms of the $\alpha + \alpha + ^{40}$Ca system  with the Generator Coordinate Method, the level $J^{\pi} = 1^{-}$ of the theoretical $K^{\pi} = 0^{-}$ band is predicted at $E_{\mathrm{th}} \approx -9.1$ MeV with respect to the $\alpha + \alpha + ^{40}$Ca threshold, i.e., $E_x \approx 3.7$ MeV. In the $^{48}$Cr experimental spectrum \cite{ENSDF}, there are not negative parity levels with defined assignments for an association with the theoretical $E_x(1^{-})$ energies mentioned.

A first calculation of $^{64}$Ge was presented by M.~A.~Souza {\it et al.}~in Ref.~\cite{SMB2019_JPCS} using the \mbox{(1 + Gaussian)$\times$(W.S.~+ W.S.$^3$)} potential and the same parameters applied in the present work; the $G = 12$ and $G = 13$ bands of the $\alpha + ^{60}$Zn system were calculated, using the depths $V_0 = 220$ MeV and $V_0 = 241$ MeV, respectively. In the case of the $G = 13$ negative parity band, the increase in $V_0$ was necessary for a better reproduction of the incomplete experimental negative parity band which starts at the $J^{\pi} = (3^{-})$ level with $E_x = 2.970$ MeV. By using $V_0 = 220$ MeV, the $1^{-}$ bandhead is found at $E_x = 7.542$ MeV, which is incompatible with the experimental band considered.

In Ref.~\cite{SM2015}, the present authors showed a study on the $\alpha$ + core structure in $^{90}$Sr, $^{92}$Zr, $^{94}$Mo, $^{96}$Ru, and $^{98}$Pd nuclei, using a nuclear potential of \mbox{W.S.~+ W.S.$^3$} shape; in that work, an exploratory calculation of the negative parity bands was made for $^{92}$Zr, $^{94}$Mo, $^{96}$Ru, and $^{98}$Pd, where a depth $V_0 = 238$ MeV was employed to give a satisfactory reproduction of the incomplete experimental negative parity bands, while $V_0 = 220$ MeV was applied to the ground state bands. Using the \mbox{$\alpha$ + core} potential of the present work with the fixed depth $V_0 = 220$ MeV, the $1^{-}$ bandheads for $^{92}$Zr ($G = 15$) and $^{94}$Mo ($G = 17$) are found at $E_x = 7.514$ MeV and 7.084 MeV, respectively; the $E_x(1^{-})$ value mentioned for $^{94}$Mo is similar to other predictions on the $G = 17$ band \cite{O1995,M2008,SM2005}. However, the calculated $E_x(1^{-})$ values with $V_0 = 220$ MeV are above the experimental bands considered in Ref.~\cite{SM2015}. Currently, there is not experimental data for $^{92}$Zr and $^{94}$Mo that characterize negative parity bands with bandhead at $E_x \approx 7$ MeV.

New experimental data related to the $22 \leq Z \leq 42$ region are needed to define whether the depth $V_0$ is necessarily a band dependent parameter. The calculations presented in Refs.~\cite{SM2015,IMP2019,SMB2019_JPCS} suggest that the $\alpha$ + core potential should be deeper for a satisfactory description of the negative parity bands of some nuclei in this region. The data obtained through $\alpha$-transfer reactions are especially important for the characterization of the excited bands in the context of the $\alpha$ + core structure.

\section{Summary and conclusions}

This work shows a systematic study of the $\alpha + \mathrm{core}$ structure in even-even nuclei of the $22 \leq Z \leq 42$ region using the Local Potential Model and the nuclear potential of \mbox{(1 + Gaussian)$\times$(W.S.~+ W.S.$^3$)} shape with two free parameters. This potential has already been successfully tested in nuclei of different mass regions with the same set of fixed parameters \cite{SMB2019, SM2017}. A selection criterion based on $Q_{\alpha}/A_T$ has been applied in the even-$Z$ isotopic chains from Ti to Mo, resulting in 20 selected nuclei of the $22 \leq Z \leq 42$ region, plus the $^{48}$Cr nucleus which was exceptionally included in the comparative study.

The ground state bands of the $\alpha + \mathrm{core}$ systems were calculated, producing a good general description of the experimental spectra, mainly from the $0^{+}$ state to the $8^{+}$ state. An important result on the energy spectra is that the model performed similarly both in the nuclei with the $\alpha + \mathrm{DCSC}$ configuration and the selected nuclei without such a configuration.

An analysis of the reduced $\alpha$-width, dimensionless reduced $\alpha$-width, and the ratio $\mathcal{R}$ for the $0^{+}$ ground state shows that the $^{44}$Ti and $^{94}$Mo nuclei have a more pronounced $\alpha$-cluster structure than other selected nuclei in their respective mass subregions, corroborating the statement that $^{44}$Ti and $^{94}$Mo are preferential nuclei for $\alpha$-clustering. The $^{60}$Zn nucleus has a reduced $\alpha$-width $\gamma _\alpha ^2(0^{+})$ considerably higher than in other selected nuclei of the same subregion, suggesting this nucleus is also preferential for $\alpha$-clustering. Furthermore, it is indicated there are other nuclei, with or without the $\alpha + \mathrm{DCSC}$ configuration, showing a significant $\alpha$-clustering degree compared to $^{44}$Ti, $^{60}$Zn, and $^{94}$Mo: in the $^{44}$Ti subregion, the $^{52}$Ti, $^{46}$Cr, $^{54}$Cr, and $^{48}$Cr nuclei stand out with $\gamma _\alpha ^2\,/\,\gamma _\alpha ^2(^{44}\mathrm{Ti}) = 0.531$, 0.509, 0.278, and 0.254, respectively; in the $^{60}$Zn subregion, the $^{56}$Fe, $^{58}$Fe, $^{58}$Ni, $^{60}$Ni, $^{64}$Ge, and $^{68}$Se nuclei stand out with $\gamma _\alpha ^2\,/\,\gamma _\alpha ^2(^{60}\mathrm{Zn}) = 1.015$, 0.772, 0.810, 0.635, 0.563, and 0.328, respectively; in the $^{94}$Mo subregion, the $^{82}$Zn and $^{84}$Ge nuclei stand out with $\gamma _\alpha ^2\,/\,\gamma _\alpha ^2(^{94}\mathrm{Mo}) = 0.249$ and 0.278, respectively. A complementary analysis of the results in our previous work \cite{SM2015} indicates that $^{96}$Ru and $^{98}$Pd also have an expressive $\alpha$-clustering degree compared to $^{94}$Mo.

The ratio of the calculated rms charge radius for the total nucleus ($\langle r^{2}\rangle_{T}^{1/2}$) to the respective experimental rms radius ($\langle r^{2}\rangle_{T\:\mathrm{exp}}^{1/2}$) was determined for 11 nuclei from the selected set, obtaining ratios above 0.99 in most cases. The very good agreement between calculated and experimental rms radii is demonstrated for the $\alpha$ + core model.

The $B(E2)$ rates were calculated for the $2^{+} \rightarrow 0^{+}$ and $4^{+} \rightarrow 2^{+}$ transitions of the g.s.~band. Most of the calculated values reproduce the order of magnitude of the respective experimental data without the use of effective charges. Concerning the absolute values, there is reasonable agreement between theoretical and experimental $B(E2)$ rates for $^{52}$Ti, $^{58}$Ni, $^{90}$Sr, and $^{92}$Zr.
Again, it should be noted that the relatively satisfactory results for the $B(E2)$ rates include nuclei with and without the $\alpha + \mathrm{DCSC}$ configuration. For $^{72}$Kr and $^{78}$Sr, it is indicated that the single contribution of the $\alpha + \mathrm{core}$ structure in the $2^{+} \rightarrow 0^{+}$ and $4^{+} \rightarrow 2^{+}$ transitions is not sufficient to describe the strong collectivity observed in the two mentioned nuclei, unless by applying substantial effective charges. Despite the discrepancy between calc.~and expt.~$B(E2)$ values for $^{72}$Kr, the calculated reduced $\alpha$-width \mbox{$\gamma _\alpha ^2$($G_{\mathrm{g.s.}} = 14;0^{+}$)} suggests this nucleus has a relatively high $\alpha$-clustering degree in its mass subregion.

It has been observed that the radial parameter $R$ has a clear linear trend with $A_T^{1/3}$ and $A_{\mathrm{core}}^{1/3}$, which has been found locally ($22 \leq Z \leq 42$) and in nuclei of different mass regions by the set \{$^{20}$Ne, $^{44}$Ti, $^{94}$Mo, $^{212}$Po\}. A study of the free parameter $\sigma$ in the $22 \leq Z \leq 42$ region indicates a linear trend of $\sigma$ with the $0^{+}$ state energy predicted by the \mbox{W.S.~+ W.S.$^3$} nuclear potential.

In conclusion, the Local Potential Model with the \mbox{(1 + Gaussian)$\times$(W.S.~+ W.S.$^3$)} nuclear potential shows its comprehensiveness in describing the properties of the g.s.~bands of $22 \leq Z \leq 42$ even-even nuclei in terms of an $\alpha + \mathrm{core}$ system, providing a good account of the experimental data in general and a systematics applicable even in nuclei without the $\alpha + \mathrm{DCSC}$ configuration.
In addition to indicating that $^{44}$Ti and $^{94}$Mo are preferential nuclei for $\alpha$-clustering in their respective mass subregions, the model points the existence of nuclei without the $\alpha + \mathrm{DCSC}$ configuration in the $22 \leq Z \leq 42$ region with an expressive $\alpha$-clustering degree compared to $^{44}$Ti, $^{60}$Zn, and $^{94}$Mo. Applications of the model in other regions, such as $Z < 22$ and $Z > 42$, the extension of the calculations for the $\alpha$ + core excited bands and odd nuclei are possibilities for future work.

\begin{acknowledgments}
The authors thank the HPC resources provided by Information Technology Superintendence (HPC-STI) of University of S\~{a}o Paulo.
Support from Instituto Nacional de Ci\^{e}ncia e Tecnologia -- F\'{\i}sica Nuclear e Aplica\c{c}\~{o}es (INCT-FNA) is acknowledged.
\end{acknowledgments}

\end{document}